\documentclass[11pt]{article}

\usepackage[margin=1in]{geometry}
\usepackage{times}
\usepackage[T1]{fontenc}
\usepackage[utf8]{inputenc}

\usepackage{authblk}
\usepackage{float}
\usepackage{graphicx}
\usepackage{subcaption}
\usepackage{amsmath}
\usepackage{amssymb}
\usepackage{amsfonts}
\usepackage{amsthm}
\usepackage{algorithm}
\usepackage[noend]{algpseudocode}
\usepackage{enumitem}
\usepackage{multirow}
\usepackage{array}
\usepackage{textcomp}
\usepackage[table]{xcolor}
\usepackage{colortbl}
\usepackage{tikz}
\usetikzlibrary{positioning,arrows.meta,shapes.geometric}
\usepackage[normalem]{ulem}
\usepackage{listings}
\usepackage{booktabs}
\usepackage{xspace}
\usepackage[numbers,sort&compress]{natbib}
\usepackage{url}
\usepackage{adjustbox}
\usepackage{hyperref}

\newcommand{\TetraRL}{\mbox{\textsc{TetraRL}}\xspace}
\newcommand{\Rfour}{R$^4$\xspace}
\newcommand{\Rthree}{R$^3$\xspace}

\newcommand{\takeaway}[2]{
\par\vspace{2.5mm}\noindent
\begin{tikzpicture}
\node[
 draw=black, line width=0.9pt, fill=white,
 rounded corners=1.2pt,
 inner xsep=6pt, inner ysep=6pt,
 text width=\dimexpr0.97\columnwidth-12pt\relax,
 align=left, font=\small
] (tkbody) {\rule[-0.3ex]{0pt}{3.4ex}\ignorespaces #2};
\node[
 fill=black, text=white, font=\footnotesize\bfseries,
 inner xsep=6pt, inner ysep=2.5pt,
 anchor=south west
] (tkhead) at ([xshift=4pt,yshift=0.3pt]tkbody.north west) {\strut #1};
\end{tikzpicture}\par\vspace{1.5mm}
}

\begin{document}

\title{\TetraRL: A Self-Adaptive Runtime for On-Device Deep Reinforcement Learning Systems}

\author[1]{Zexin Li}
\author[1]{Soheil Shirvani}
\author[1]{Cong Liu}
\affil[1]{University of California, Riverside}
\affil[ ]{\normalsize\texttt{\{zli536, sshir009, congl\}@ucr.edu}}

\date{}

\maketitle

\begin{abstract}
Autonomous robotic systems, such as autonomous vehicles, drones, and mobile robots, increasingly require on-device Deep Reinforcement Learning (DRL) to continuously adapt to dynamic environments. Unlike cloud-based learning, embedded DRL must perform training and inference directly on resource-constrained hardware while maintaining timely decision-making. This requirement exposes a fundamental challenge: on-device DRL must simultaneously balance four tightly coupled objectives: real-time performance, task reward, memory utilization, and energy consumption. Optimizing these objectives independently often leads to suboptimal system behavior, while naïve multi-objective optimization can violate resource constraints and degrade reliability.

This paper presents \TetraRL{}\footnote{The name \TetraRL{} derives from the Greek prefix ``tetra-'', meaning ``four'', reflecting the four optimization objectives in the \Rfour{} principle: real-time, reward, RAM, and reserve.}, a holistic runtime framework for self-adaptive tetra-objective on-device DRL. \TetraRL{} formulates embedded DRL as a unified optimization problem over real-time, reward, RAM, and reserve (energy) objectives, and employs a preference-conditioned reinforcement learning controller to dynamically navigate the resulting trade-off space. The framework further integrates a unified resource-management abstraction, hardware-aware DVFS control, and a runtime Override Layer for enforcing resource constraints. We implement and evaluate \TetraRL{} across diverse DRL environments and embedded platforms, including NVIDIA Jetson AGX Orin and Orin Nano. Experimental results demonstrate that \TetraRL{} consistently auto-balances the four objectives, achieving competitive trade-offs across them while maintaining negligible runtime overhead. Furthermore, \TetraRL{} enables runtime-switchable optimization goals through a single trained policy, providing a practical foundation for self-adaptive and resource-aware on-device DRL.
\end{abstract}


\section{Introduction}
\label{sec:intro}

Deep Reinforcement Learning (DRL) has emerged as a powerful paradigm for enabling autonomous decision making in complex and dynamic environments~\cite{mnih2015human,schulman2017proximal}. Recent advances have accelerated the deployment of DRL on resource-constrained edge platforms, including autonomous robots, drones, and assisted-driving systems~\cite{kahn2018selfsupervised,kiran2021deepdrive,aggravi2021haptic}. Unlike cloud-centric learning pipelines, these systems increasingly require \emph{on-device} adaptation, where the agent continuously learns and retrains directly on embedded hardware while simultaneously performing runtime inference. Such capability is critical in real-world deployments where environmental conditions evolve rapidly, and communication with cloud infrastructure may be unavailable, unreliable, or prohibitively expensive.

For example, autonomous mobile delivery robots operating in unfamiliar environments must continuously adapt to changing terrain, obstacles, and mission objectives~\cite{kahn2018selfsupervised,aggravi2021haptic}. Similarly, autonomous vehicles must retrain their decision-making policies to accommodate evolving traffic patterns, weather conditions, and sensor characteristics~\cite{kiran2021deepdrive}. In both cases, the DRL agent must learn from newly collected experiences while maintaining timely decision-making and respecting the strict resource limitations of embedded platforms. These examples highlight the growing need for efficient on-device DRL training that can simultaneously satisfy real-time constraints, maintain algorithmic performance, and operate within limited hardware resources.

\begin{figure}[!tbp]
\centering
\includegraphics[width=0.7\textwidth]{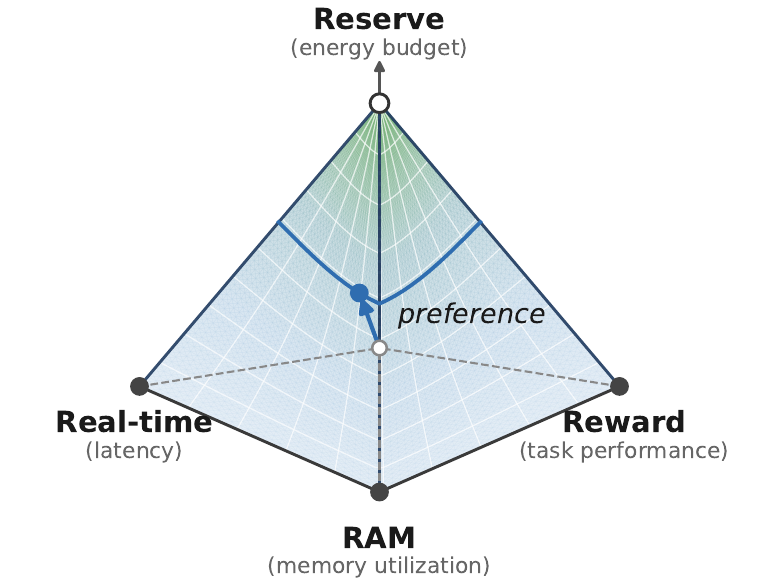}
\caption{Optimization challenges of on-device DRL. Embedded DRL systems must jointly balance four frequently conflicting objectives: $R_1$ real-time performance (latency / timeliness), $R_2$ reward, $R_3$ memory utilization, and $R_4$ energy consumption. Improving one objective often degrades another, creating a complex four-dimensional optimization space.}
\label{fig:r4_simplex}
\vspace{-3mm}
\end{figure}

However, conducting on-device DRL training remains challenging. As illustrated in Figure~\ref{fig:r4_simplex}, embedded DRL systems must simultaneously optimize four tightly coupled and often conflicting objectives:

\begin{itemize}[leftmargin=10pt]
    \item \textbf{Real-time performance ($R_1$):} robotic control loops require predictable and low execution latency.
    \item \textbf{Reward ($R_2$):} the learning algorithm must maintain high task performance and adaptation quality.
    \item \textbf{Memory utilization ($R_3$):} DRL training competes for limited unified CPU/GPU memory with perception, planning, and operating-system services.
    \item \textbf{Energy consumption ($R_4$):} battery-powered embedded platforms must operate under strict power and thermal budgets.
\end{itemize}

These objectives are inherently interconnected. Increasing training batch size may improve hardware utilization but can increase latency and memory pressure. Enlarging replay buffers may improve learning quality but consumes additional memory resources. Raising processor frequencies through Dynamic Voltage and Frequency Scaling (DVFS) can reduce execution time but increases energy consumption and thermal stress. Conversely, aggressive energy-saving configurations may prolong training and degrade responsiveness. Optimizing any single objective independently, therefore often results in suboptimal system behavior, while naively optimizing multiple objectives may violate hard resource constraints or even trigger system failures.

Existing approaches each cover only a slice of this four-objective space, and the missing slices are not optional. \Rthree~\cite{li2023r3} controls latency and memory through adaptive batch-size and replay-buffer management but ignores energy, so its configurations can silently drain a battery-powered platform that must also respect an energy-reserve budget. DuoJoule~\cite{duojoule2024} co-optimizes latency and energy through DVFS but does not bound the unified CPU/GPU memory that on-device training shares with perception and planning, so its configurations can trigger out-of-memory aborts on memory-constrained boards such as the Orin Nano. Crucially, both treat \emph{reward as fixed}: they tune system knobs around an unchanged training schedule and never reason about the latency-reward or energy-reward trade-off, even though on-device training can spend or save large amounts of compute by adapting how long and how hard it trains. From the algorithm side, mainstream DRL methods~\cite{mnih2015dqn,vanhasselt2016double,bellemare2017distributional,mnih2016asynchronous,schulman2017ppo} optimize reward alone and expose no interface for latency, memory, or energy at all. The gap is therefore not a missing feature but a missing joint formulation: no existing framework manages real-time performance, reward quality, memory utilization, and energy consumption together, nor lets a deployed agent re-weight these objectives at runtime as battery, deadline, and memory pressure change.

To address this challenge, we introduce \TetraRL{}, a self-adaptive runtime framework for on-device deep reinforcement learning. We formulate embedded DRL as a four-dimensional optimization problem over \emph{real-time}, \emph{reward}, \emph{RAM} (memory), and \emph{reserve} (energy)
objectives, collectively referred to as the \textbf{\Rfour} principle. Rather than treating resource constraints as fixed design-time assumptions, \TetraRL{} learns a preference-conditioned policy that continuously adapts system behavior according to runtime optimization goals. By embedding user preferences directly into the policy state space, a single trained policy can dynamically auto-balance the four objectives and switch between operating points without retraining.

The design of \TetraRL{} consists of four key components. First, a preference-conditioned controller jointly reasons about the trade-offs among the four objectives. Second, a unified resource-primitive abstraction enables consistent management of both off-policy and on-policy DRL algorithms. Third, a hardware-aware DVFS management layer exposes power-performance trade-offs as controllable actions. Finally, an Override Layer enforces resource constraints at runtime and prevents excessive violations that may occur under purely learning-based optimization. Together, these components enable \TetraRL{} to achieve adaptive multi-objective optimization while preserving practical deployability on embedded systems.

We evaluate \TetraRL{} across diverse DRL environments and embedded platforms, including classic control tasks, Atari games, multi-objective reinforcement learning benchmarks, and a synthetic four-dimensional DAG-scheduling environment. Experiments are conducted on NVIDIA Jetson AGX Orin and Orin Nano platforms under realistic resource constraints. The results demonstrate that \TetraRL{} consistently auto-balances the four objectives, delivering competitive trade-offs across them while maintaining low runtime overhead and robust adaptation to changing system conditions.

The major contributions of this paper are summarized as follows:

\begin{itemize}[leftmargin=10pt]
    \item \textbf{The first unified \Rfour framework for on-device DRL.} We formulate embedded DRL as a four-objective optimization problem spanning real-time performance, reward quality, memory utilization, and energy consumption, and present \TetraRL{}, the first runtime framework that jointly manages all four objectives.
    
    \item \textbf{Self-adaptive optimization with strong latency--reward trade-offs.} We develop a preference-conditioned DRL framework that auto-balances the four objectives and enables runtime-switchable optimization goals through a single policy without retraining. 
    
    \item \textbf{Broader support for both off-policy and on-policy DRL.} Through a unified resource-primitive abstraction, \TetraRL{} extends resource-aware optimization beyond replay-buffer-based algorithms and provides a common runtime management interface across five heterogeneous DRL methods.
    
    \item \textbf{Comprehensive evaluation on real embedded platforms with negligible overhead.} We conduct extensive experiments across multiple DRL environments, optimization objectives, and two Jetson platforms. On the budget-matched protocol \TetraRL{} tracks the latency-aware \Rthree{} controller to within $1\%$ on latency, and its in-loop control adds only $0.078$\,ms ($4.4\%$) of pure-framework per-step latency overhead and under $1\%$ of process memory, confirming that the framework can be deployed essentially for free on top of the bare RL pipeline. Against representative resource-control baselines (\Rthree~\cite{li2023r3}, DuoJoule~\cite{duojoule2024}, and the MAX-A/MAX-P frequency-pinning schedules), \TetraRL{} cuts time-to-solve latency (wall-clock latency to first reach the target reward)
by $53.6$--$74.6\%$ on Atari-Breakout/Orin~Nano, and on CartPole-v0/Orin~AGX reaches the same reward target on a substantially smaller data budget rather than running the fixed schedule faster, while preserving competitive reward (e.g.\ DQN Breakout reward $320.0$ vs.\ MAX-A's $250.0$ on AGX, and $+26.7\%$ to $+334.0\%$ reward over \Rthree{} on Nano). These gains reflect that \TetraRL{}'s preference-conditioned early-stop reaches the same reward target on a smaller data budget rather than running the same fixed schedule faster; under a matched fixed step budget, the per-step overhead is the negligible $4.4\%$ reported above.
\end{itemize}

\noindent\textbf{Differences from prior conference papers.}
This journal article substantially extends our prior conference publications, \Rthree~\cite{li2023r3} and DuoJoule~\cite{duojoule2024}. Compared with \Rthree, which focuses on latency and memory management, and DuoJoule, which focuses on latency-energy co-optimization, \TetraRL{} introduces a unified four-objective optimization framework, preference-conditioned policy learning, runtime-switchable multi-objective auto-balancing, hardware-aware DVFS control, support for both off-policy and on-policy DRL algorithms, and a significantly expanded evaluation methodology covering multi-objective benchmarks, dynamic preference switching, DVFS characterization, and extensive cross-platform experiments.

\section{Background}
\label{sec:background}

Deep Reinforcement Learning (DRL) trains an agent to maximize the expected discounted return by interacting with an environment, and underpins modern autonomous decision making in robotics, drones, and assisted-driving systems~\cite{mnih2015human,schulman2017proximal,mnih2015dqn}. While inference is increasingly pushed to the edge, a growing class of deployments also requires \emph{on-device training}: when an autonomous robot or vehicle encounters terrain, obstacles, traffic, or sensor characteristics that drift away from its pretraining distribution, it must keep adapting its policy locally because cloud connectivity is often unavailable, unreliable, or too costly to be in the control loop~\cite{kahn2018selfsupervised,kiran2021deepdrive,aggravi2021haptic}. Doing this directly on embedded hardware means the same platform must sustain timely decision making while running the learning algorithm under tight memory and power budgets. The levers available to manage this regime are the resource and frequency knobs already present on these platforms: on the algorithm side, the replay/rollout configuration (e.g.\ minibatch size, replay capacity, and update schedule), and on the hardware side, Dynamic Voltage and Frequency Scaling (DVFS) over CPU and GPU operating frequencies, which trades execution time against energy and thermal headroom~\cite{duojoule2024,dvfsdrl2024}.

We evaluate four quantities, all measured from real on-device telemetry. \emph{Latency} $L$ is the realized per-step or per-update wall-clock time, the timeliness proxy for the real-time objective. \emph{Memory} $M$ is peak unified CPU/GPU memory utilization during training, reported as a fraction of total device memory, since DRL training competes for the same unified memory as perception, planning, and OS services. \emph{Maximal reward} $R_{\max}$ is the best task return achieved over a run (e.g.\ raw game score on Atari-Breakout), capturing learning quality independent of the system overhead. Because $L$ and $M$ are costs (lower is better) while $R_{\max}$ is a utility (higher is better), we normalize each axis and orient them consistently before any cross-method comparison.

\section{Motivation}
\label{sec:motivation} 

\begin{table}[!htbp]
\centering
\caption{Workload drift within a single on-device training run motivates soft resource objectives over static hard caps. Both peak unified-memory utilization (a) and mean power draw (b) for DQN on Atari-Breakout, measured on an NVIDIA Jetson AGX Orin (32\,GB unified memory; $200$\,k environment steps, single seed, $100$\,k replay capacity, gradient updates begin at step $50$\,k), bucketed in $10$\,k-step windows. Memory drifts $2.69\times$ as the off-policy replay buffer first warms and then fills, while power draw drifts $1.40\times$ once gradient updates begin. This substantial intra-run drift in both resources shows that a static cap is either overly restrictive early on or ineffective after convergence.}
\label{tab:workload_drift}
\renewcommand{\arraystretch}{1.15}
\begin{subtable}[t]{0.48\linewidth}
\centering
\caption{Memory drift}
\label{tab:workload_drift_mem}
\setlength{\tabcolsep}{3pt}
\resizebox{\linewidth}{!}{
\begin{tabular}{l c c}
\toprule
Phase & Step window & Peak memory\\
\midrule
\multirow{2}{*}{Warm-up} & $0$--$10$\,k & \textbf{$10.4\%$}\\
 & $40$--$50$\,k & $17.4\%$\\
\midrule
\multirow{2}{*}{Training} & $50$--$60$\,k & $20.3\%$\\
 & $90$--$100$\,k & $27.4\%$\\
\midrule
\multirow{2}{*}{Plateau} & $130$--$140$\,k & $27.9\%$\\
 & $190$--$200$\,k & \textbf{$27.9\%$}\\
\midrule
\multicolumn{2}{l}{Intra-run drift} & $\mathbf{2.69\times}$\\
\bottomrule
\end{tabular}}
\end{subtable}
\hfill
\begin{subtable}[t]{0.48\linewidth}
\centering
\caption{Energy drift}
\label{tab:workload_drift_energy}
\setlength{\tabcolsep}{3pt}
\resizebox{\linewidth}{!}{
\begin{tabular}{l c c}
\toprule
Phase & Step window & Mean power\\
\midrule
\multirow{2}{*}{Warm-up} & $0$--$10$\,k & \textbf{$4.6$\,W}\\
 & $40$--$50$\,k & $4.7$\,W\\
\midrule
\multirow{2}{*}{Training} & $50$--$60$\,k & $6.3$\,W\\
 & $90$--$100$\,k & $6.4$\,W\\
\midrule
\multirow{2}{*}{Plateau} & $130$--$140$\,k & $6.4$\,W\\
 & $160$--$170$\,k & \textbf{$6.4$\,W}\\
\midrule
\multicolumn{2}{l}{Intra-run drift} & $\mathbf{1.40\times}$\\
\bottomrule
\end{tabular}}
\end{subtable}
\end{table}

\subsection{Why resource budgets must be soft objectives, not hard caps}
\label{sec:mot:objectives}

\begin{table}[!htbp]
\centering
\caption{Hard-budget infeasibility for both memory and power. Sweeping a static cap over the same DQN-Breakout $200$\,k run with hard-abort-on-cap. (a) Static unified-memory caps swept $12$--$40\%$: every cap $\le 28\%$ aborts mid-training, while caps $\ge 32\%$ complete the full $200$\,k steps without ever binding. (b) Static power caps swept over the same run on real on-device telemetry (MAXN, jetson\_clocks): every cap at or below the converged training draw forces a hard abort, while a cap above the peak draw never binds. In neither resource is any single static cap both feasible during warm-up and constraining at convergence.}
\label{tab:hard_budget_infeasibility}
\renewcommand{\arraystretch}{1.15}

\begin{subtable}[t]{0.3\linewidth}
\centering
\caption{Memory budget}
\label{tab:hard_budget_infeasibility_mem}
\setlength{\tabcolsep}{3pt}
\resizebox{\linewidth}{!}{
\begin{tabular}{c c c}
\toprule
Static cap & Abort step & Outcome\\
\midrule
$12\%$ & ${\sim}19.9$\,k & Aborts (warming)\\
$16\%$ & ${\sim}42.8$\,k & Aborts\\
$20\%$ & ${\sim}57.7$\,k & Aborts\\
$24\%$ & ${\sim}80.4$\,k & Aborts\\
$28\%$ & ${\sim}137.8$\,k & Aborts (near peak)\\
\midrule
$32\%$ & --- & Full $200$\,k (never binds)\\
$40\%$ & --- & Full $200$\,k (never binds)\\
\bottomrule
\end{tabular}}
\end{subtable}
\hfill
\begin{subtable}[t]{0.66\linewidth}
\centering
\caption{Power budget}
\label{tab:hard_budget_infeasibility_power}
\setlength{\tabcolsep}{3pt}
\resizebox{\linewidth}{!}{
\begin{tabular}{c c c c c c}
\toprule
Power cap & Abort step & Max reward & Energy & Mean power & Outcome\\
\midrule
$4.0$\,W & ${\sim}0.0$\,k & --- & $0.0$\,kJ & $4.3$\,W & Aborts\\
$4.5$\,W & ${\sim}9.4$\,k & $0.26$ & $0.1$\,kJ & $4.4$\,W & Aborts\\
$5.0$\,W & ${\sim}50.2$\,k & $0.35$ & $0.3$\,kJ & $4.6$\,W & Aborts\\
$6.0$\,W & ${\sim}51.0$\,k & $0.35$ & $0.4$\,kJ & $4.6$\,W & Aborts\\
\midrule
$7.0$\,W & --- & $2.60$ & $8.8$\,kJ & $5.9$\,W & Full $200$\,k (never binds)\\
\bottomrule
\end{tabular}}
\vspace{-3mm}
\end{subtable}

\end{table}
\Rthree~\cite{li2023r3} and DuoJoule~\cite{duojoule2024}, the two closest prior systems on Jetson hardware, both target on-device DRL \emph{training} (not merely inference), yet both treat their critical resource as a \emph{hard constraint} guarded by a hand-tuned controller: \Rthree gates batch size against a latency-driven moving-average tracker, and DuoJoule's MetricTracker collapses energy and latency into a single discrete efficiency-mode switch. While a static budget is reasonable when the workload is stationary, on-device training is non-stationary, and three compounding reasons make a hard cap ill-suited here.

First, the memory footprint is non-stationary by construction: it grows over the course of a run as the replay buffer warms and fills, rather than holding at the fixed level a static cap assumes. This growth is driven primarily by the off-policy DQN replay buffer, which fills from warm-up to its configured capacity over training and is the dominant consumer of the rising peak unified-memory footprint. In a representative run on an NVIDIA Jetson AGX Orin (32\,GB unified memory), DQN on Atari-Breakout, 200\,k environment steps, single seed, with the same $100$\,k replay capacity used in our on-device evaluation, the peak unified-memory utilization drifts by $2.69\times$ (from $10.4\%$ to $27.9\%$, i.e.\ from about $3.3$\,GB to $8.9$\,GB on the $32$\,GB board) between the first and last $10\,$k training steps as the replay buffer fills (Table~\ref{tab:workload_drift}). The same run's power draw drifts similarly, rising from about $4.6$\,W during warm-up to about $6.4$\,W ($1.40\times$) once gradient updates begin at step $50$\,k (Table~\ref{tab:workload_drift_energy}). A hard ceiling chosen at deployment time is therefore either (i) below the eventual steady-state footprint, in which case the run may proceed through warm-up but aborts later as the replay buffer fills and gradient updates begin, or (ii) above the peak footprint to guarantee completion, in which case it never constrains the run and provides no useful steering signal. Table~\ref{tab:hard_budget_infeasibility} shows this directly: memory caps at or below the converged peak eventually abort, while caps above the peak complete without ever binding.   We make this concrete in Table~\ref{tab:hard_budget_infeasibility}: sweeping a static unified-memory cap from $12\%$ to $40\%$ over the same $200$\,k DQN-Breakout run with an R\textsuperscript{3}-style hard-abort-on-cap, every cap at or below the $27.9\%$ converged peak forces a hard abort partway through training, a $12\%$ cap aborts at step ${\sim}19.9$\,k while the replay buffer is still warming, and even a $28\%$ cap (essentially at the converged peak) aborts at step ${\sim}137.8$\,k, while every cap strictly above the peak ($\ge 32\%$) runs the full $200$\,k steps without ever binding, leaving no value that is simultaneously feasible during warm-up and constraining at convergence. This infeasibility is precisely why \TetraRL{} adopts a preference-conditioned PPO arbiter that adapts the effective budget online rather than committing to a static ceiling chosen at deployment time.
 
Second, hard-constraint violations in DRL training are not recoverable in the same way as in inference. A missed latency target in an inference pipeline drops one frame; a memory-cap or energy-cap abort in training can instead terminate the process or discard a partially-applied update, stalling or destabilizing the run rather than dropping a single step~\cite{mnih2015dqn,lillicrap2015ddpg}. Penalizing budget overruns softly keeps the learning signal alive while still steering the policy away from the budget, and a separate hardware-level safety guardrail can catch the rare hard violations as a last-resort backup, so the learner never has to choose between staying within budget and continuing to learn.

Third, the user-facing semantics are themselves soft, and energy is where this matters most. On-device DRL training is interesting precisely because it lets an agent keep adapting after deployment, but the deployments that need this are exactly the ones running untethered on a fixed energy supply: a field robot or drone that refines its policy between charges, a battery- or solar-powered sensor node that must keep learning for days on a single charge, or a wearable that adapts on the user's device without draining it. In all of these, energy is not a tidy hard ceiling but a budget that the deployment context keeps changing, battery state of charge falls over a mission, a harvesting node's input varies with sunlight, and a platform's sustainable power envelope is itself a limit. That envelope is concrete: each board has an instantaneous power level it can deliver and dissipate before it throttles, and compute-heavy update steps can spike draw past it, at which point the platform caps its clocks (slowing training) or, in the worst case, browns out the device. A robot whose battery has dropped to $30\%$ should therefore be \emph{biased} toward energy efficiency, not forbidden from a control action that would briefly push power high, and a node running low on harvested energy should ease off rather than abort; a tightening end-of-episode latency target should likewise steer the policy without hard-clipping it.

These preferences should be tunable at runtime, and serving them should not demand heavy extra machinery, a separate hand-tuned controller per regime, or offline profiling for every new battery or platform. The gap, then, is the absence of a single mechanism that exposes energy alongside the other resources as a soft, tunable preference the learner can be steered by online, so the same trained agent can switch between, e.g., a battery-saver and a performance regime as conditions change rather than being locked to one operating point chosen at deployment time.

\takeaway{Takeaway: Soft budgets, not hard caps}{On-device training shifts as it learns, so any fixed memory or energy limit is either too tight and kills the run early or too loose and never bites. All four resource dimensions should be soft, tunable, optimizable  objectives the runtime steers toward, not rigid limits.}

\subsection{Why 4-D co-optimization (\Rfour) cannot be decomposed}
\label{sec:mot:fourd}

\begin{figure}[!htbp]
\centering

\begin{subfigure}[t]{0.48\textwidth}
    \centering
    \includegraphics[width=\linewidth]{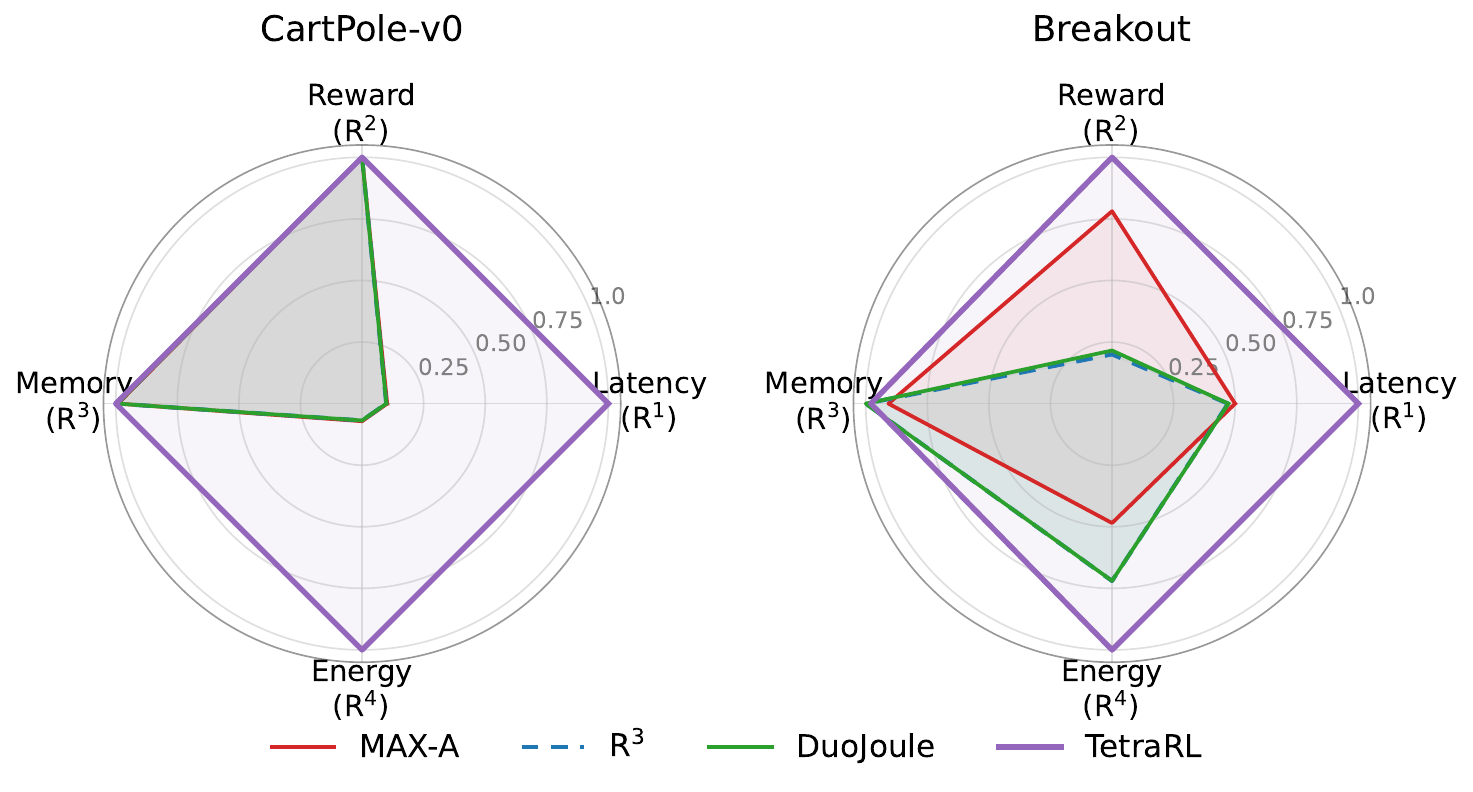}
    \caption{Controller coverage across two workloads on Orin AGX}
    \label{fig:motivation_radar_workload}
\end{subfigure}
\hfill
\begin{subfigure}[t]{0.48\textwidth}
    \centering
    \includegraphics[width=\linewidth]{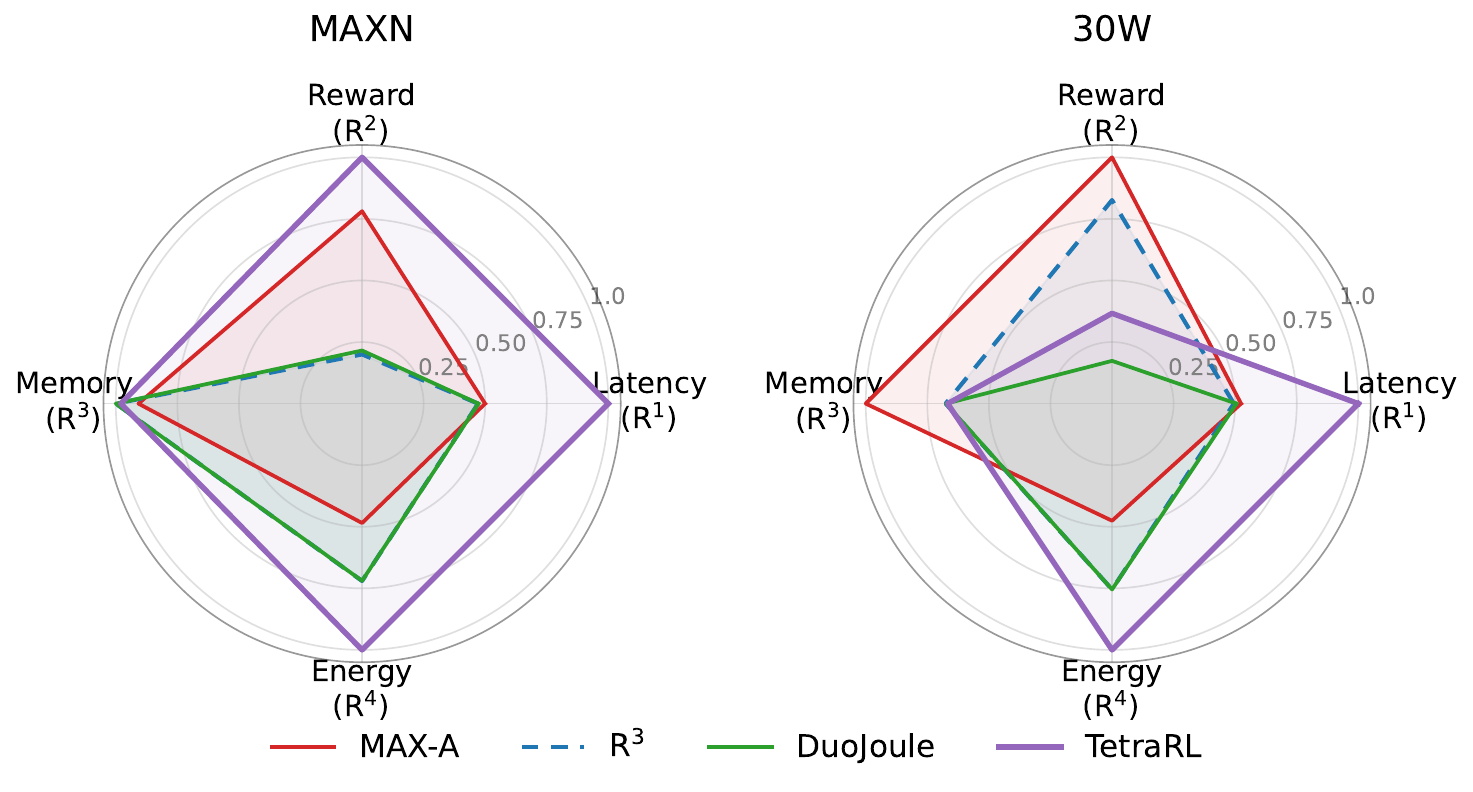}
    \caption{Controller coverage under two power conditions on Orin AGX }
    \label{fig:motivation_radar_power}
\end{subfigure}

\caption{Empirical \Rfour{} coverage of four controllers under changing operating conditions. (a) Coverage across two workloads on the same hardware platform. (b) Coverage under two power envelopes on the same hardware platform. In all radar plots, each axis is normalized so that the outer ring corresponds to the per-axis best value across controllers. Changes in workload or power budget substantially alter controller rankings and reduce the coverage of lower-dimensional heuristics, whereas \TetraRL{} consistently remains at or near the outer ring across conditions.}
\label{fig:motivation_radar}
\end{figure}

Once memory and energy are soft objectives rather than hard caps, the natural next question is whether they can be optimized separately, one controller per resource. They cannot: on the unified-memory edge platforms we target, the objectives interact through shared hardware knobs and a shared memory--power substrate. Raising the GPU DVFS frequency shortens per-step latency but drives energy up super-linearly, and sustained high DRAM occupancy raises the refresh and bandwidth power floor, coupling memory pressure directly to the energy reserve in a way that has no analogue on discrete-GPU hosts. An independent per-objective controller would therefore fight itself, with each controller's setpoint silently shifting cost onto the resources it does not see; this is exactly why \TetraRL{} treats the resources jointly under a single preference-conditioned arbiter rather than as decoupled per-objective loops. Table~\ref{fig:motivation_pairwise} makes these couplings concrete with real on-device sweeps: sweeping GPU DVFS frequency traces a non-monotone latency--energy frontier, and sweeping replay-buffer capacity simultaneously moves memory against both reward and energy, so no single axis can be tuned without displacing the others.

\begin{table}[!htbp]
\centering
\caption{Pairwise resource trade-offs on DQN on Atari-Breakout on AGX Orin, shown as three sub-tables. {(a)} Latency vs.\ energy as the GPU DVFS frequency is swept across its operating points ($306$--$1300$\,MHz), tracing a non-monotone frontier (minimum energy at $918$\,MHz). {(b,c)} A single replay-buffer capacity sweep ($25$k--$200$k): peak unified-memory utilization against maximal reward, and against total energy. Raising capacity past $\sim\!50$k keeps consuming memory and energy without improving reward, while shrinking it below $50$k collapses reward (and energy, since the run no longer fills). Every entry is a real on-device run; the knobs that improve one axis displace another, so the four resources cannot be optimized in isolation.}
\label{fig:motivation_pairwise}
\setlength{\tabcolsep}{4pt}
\renewcommand{\arraystretch}{1.1}
\footnotesize
\begin{minipage}[t]{0.34\linewidth}
\centering
{\footnotesize (a) Latency--energy: GPU DVFS sweep}\\[3pt]
\begin{tabular}{@{}rcc@{}}
\toprule
GPU freq.\ & Step lat.\ & Total \\
(MHz) & (ms) & energy (J) \\
\midrule
$306$  & $9.17$ & $11217.7$ \\
$510$  & $7.54$ & $9382.7$ \\
$714$  & $7.02$ & $8757.0$ \\
$918$  & $6.89$ & $8597.2$ \\
$1122$ & $6.95$ & $8687.9$ \\
$1300$ & $7.03$ & $8807.9$ \\
\bottomrule
\end{tabular}
\end{minipage}\hfill
\begin{minipage}[t]{0.32\linewidth}
\centering
{\footnotesize (b) Replay sweep: memory--reward}\\[3pt]
\begin{tabular}{@{}rcc@{}}
\toprule
Replay & Peak mem.\ & Max \\
capacity & util. & reward \\
\midrule
$25$k  & $0.153$ & $0.35$ \\
$50$k  & $0.209$ & $2.46$ \\
$100$k & $0.297$ & $2.23$ \\
$150$k & $0.385$ & $2.45$ \\
$200$k & $0.472$ & $2.42$ \\
\bottomrule
\end{tabular}
\end{minipage}\hfill
\begin{minipage}[t]{0.32\linewidth}
\centering
{\footnotesize (c) Replay sweep: memory--energy}\\[3pt]
\begin{tabular}{@{}rcc@{}}
\toprule
Replay & Peak mem.\ & Total \\
capacity & util. & energy (J) \\
\midrule
$25$k  & $0.153$ & $1693.8$ \\
$50$k  & $0.209$ & $8683.6$ \\
$100$k & $0.297$ & $8626.3$ \\
$150$k & $0.385$ & $8566.1$ \\
$200$k & $0.472$ & $8536.6$ \\
\bottomrule
\end{tabular}
\end{minipage}
\vspace{-4mm}
\end{table}

A natural counter-proposal to the 4-D framing is to stack four independent single-axis controllers, or to optimize the most coupled pair (latency--energy, as DuoJoule does) and treat the others as feed-forward inputs. We do not adopt this design for two empirical reasons rooted in the unified-memory architecture common to many edge and embedded platforms.

The first is that the four \Rfour axes are pairwise non-separable. Increasing the PPO batch size raises reward (more on-policy samples per update) but lengthens per-step latency; raising GPU DVFS frequency shortens latency but increases energy super-linearly~\cite{zhang2023dvfo,sparsedvfs2025}; shrinking the replay buffer to recover RAM cuts sample efficiency and therefore reward~\cite{horgan2018apex}; on a unified-memory device, sustained high DRAM occupancy raises the refresh and bandwidth power floor, coupling RAM directly to the energy reserve in a way that does not occur on discrete-GPU hosts. Any controller that optimizes a strict subset of these four axes silently shifts cost onto the remaining ones; this is the regime in which \Rthree (latency and RAM) and DuoJoule (latency and energy) operate, each covering only a 2-D slice of the \Rfour space and leaving the other two axes uncontrolled, the regime that the \Rfour-axis coverage table (Table~\ref{tab:axis_coverage}) makes explicit.

The second is that the best operating points across the four axes do not combine independently: the operating point that is best when one R-dimension matters most need not be best when a different dimension is prioritized, and the collection of operating points needed to serve the full range of deployment preferences is larger than simply pooling the best point for each axis in isolation. A handful of single-axis controllers therefore cannot cover the deployments that care about several R-dimensions at once, which is what a single preference-conditioned controller is meant to do. The trade-off here favors broad coverage: such a controller may give up ground to an axis-specialist when a deployment cares almost entirely about one dimension, which is the natural cost of solving the broader four-way problem rather than a single axis. The 4-D framing is therefore the minimum dimensionality at which the trade-off can be \emph{exposed} to the user as a runtime knob; lower-dimensional framings collapse the knob silently into a single hand-tuned operating point.

\takeaway{Takeaway: Co-optimize all dimensions together}{The four resource dimensions push against each other and their best settings do not combine independently, so tuning them one at a time just shifts cost onto the others. They must be optimized jointly as a single four-way problem.}

\subsection{Why an expert-based heuristic controller is insufficient}
\label{sec:mot:heuristic}

Expert-based heuristic controllers~\cite{li2023r3,duojoule2024} are a sound design for the settings they target: by treating their critical resource as a hard constraint and scoping to deployments where not all four \Rfour axes are simultaneously active, they relax the problem to a well-posed 2-D slice and solve it efficiently. The question this subsection addresses is therefore not whether such heuristics are useful, but what the full 4-D, preference-conditioned setting additionally requires when all four axes matter at once, and a lookup table indexed by the preference vector would have to stand in for a learned controller. Even granting the heuristic framing in steady state, three orthogonal regimes that on-device DRL inhabits simultaneously expose requirements beyond what a static threshold rule was designed to meet.

\begin{table}[!htbp]
\centering
\caption{Trade-off region covered by each system. \Rthree and DuoJoule each optimize a 2-D slice of the \Rfour space and treat the other two axes as fixed deployment-time choices; PD-MORL optimizes a multi-objective reward but does not specifically consider hardware (no $R_3$/$R_4$); only \TetraRL{} exposes a runtime-switchable preference vector over the full 4-D simplex.}
\label{tab:axis_coverage}
\renewcommand{\arraystretch}{1.15}
\scriptsize
\resizebox{0.6\columnwidth}{!}{
\begin{tabular}{l l l}
\toprule
System & Optimized axes & Uncovered slice\\
\midrule
\Rthree~\cite{li2023r3}        & $R_1$, $R_3$               & $R_2$, $R_4$\\
DuoJoule~\cite{duojoule2024}   & $R_1$, $R_4$               & $R_2$, $R_3$\\
DVFO~\cite{zhang2023dvfo}      & $R_1$, $R_4$ (inference)   & $R_2$, $R_3$, training\\
PD-MORL~\cite{basaklar2023pdmorl}& multi-objective $R_2$    & $R_1$, $R_3$, $R_4$\\
\hline
\TetraRL      & $R_1$, $R_2$, $R_3$, $R_4$ & ---\\
\bottomrule
\end{tabular}} 
\end{table}

Intuitively, the runtime must keep \emph{adapting} how strictly it enforces its energy and memory budgets: when behavior drifts over budget, it should tighten the pressure, and when it has slack, it can relax, continuously, in response to what it actually observes. A hand-set fixed threshold cannot do this; it fires the same way no matter how far the current behavior is from the budget, so it can neither tighten nor loosen as the workload changes. This adaptive, budget-tracking behavior is exactly what a learned controller buys and a static threshold rule cannot reproduce.

\noindent \textit{(i) Workload drift.} The same argument that disqualifies static hard caps in Section~\ref{sec:mot:objectives} disqualifies static heuristic thresholds. A DVFO-style utilization-threshold rule~\cite{zhang2023dvfo} that picks the smallest GPU frequency satisfying GPU utilization less than 85\% is well-calibrated for inference precisely because GPU utilization has a stationary distribution; in DRL training, the distribution shifts with the policy, and the threshold either over- or under-clocks for the majority of the run. A learned, preference-conditioned policy absorbs this drift through its value function and through periodic Lagrangian multiplier updates, neither of which a fixed-threshold rule can express.

\noindent \textit{(ii) Runtime drift.} Once the deployment preference is allowed to change inside an episode (a battery dropping below $20\%$ shifting the emphasis toward the energy budget, a tightening control-latency target shifting it toward real-time responsiveness), every threshold in a heuristic controller has to be re-derived from scratch, because the best trade-off itself has moved. A preference-conditioned policy instead takes the current preference directly as part of its input and adapts within a single policy step, whereas a heuristic controller adapts only after a fresh round of manual re-tuning. 

\noindent \textit{(iii) Platform drift.} The heuristic-controller knobs that worked on Orin AGX (Table~\ref{tab:dvfs_orin}) do not transfer to Orin Nano without re-measurement: the DVFS table is different, the unified-memory pressure point is different, and the thermal envelope is different. Each new platform, therefore, requires another round of manual tuning before the heuristic produces a usable trade-off. In contrast, the preference-conditioned framework retrains the policy using the new platform's measured action effects while preserving the same \Rfour{} optimization interface. Hardware-defined safety limits are enforced through a platform-aware runtime safeguard, allowing the controller logic to remain unchanged across devices. Heuristic controllers such as DVFO frequency schedulers~\cite{zhang2023dvfo} have to be manually re-tuned whenever the workload, runtime regime, or hardware platform changes. Each such change requires engineer-hours of re-measurement and re-calibration to stay at parity with a single automatic \TetraRL{} policy update, a cost that compounds across the workload-, runtime-, and platform-drift regimes discussed above.

\begin{table}[!htbp]
\centering
\caption{Measured DVFS transition latencies on NVIDIA Jetson Orin AGX, 32\,GB. CPU has 29 frequency points (115.2\,MHz to 2.20\,GHz); GPU has 11 frequency points (306\,MHz to 1.30\,GHz). All values in milliseconds per single transition (mean of 3 iterations per pair). Full per-pair tables are in the supplementary material.}
\label{tab:dvfs_orin}
\renewcommand{\arraystretch}{1.15}
\begin{tabular}{l c c c c}
\toprule
Domain & \# levels & \# pairs & Mean (ms) & Max (ms)\\
\midrule
CPU & 29 & 812 & \textbf{0.243} & \textbf{0.374}\\
GPU & 11 & 110 & \textbf{3.399} & \textbf{6.857}\\
\bottomrule
\end{tabular} 
\end{table}

\takeaway{Takeaway: Adaptation is necessary}{A fixed-threshold heuristic must be re-tuned by hand whenever the workload shifts, the user's preference changes mid-run, or the hardware platform changes. This motivates a learnable, preference-aware controller design choice.}

\section{Problem Formulation}
\label{sec:problem}

\subsection{Augmented State and Action Space}

We model \TetraRL{} as a runtime-control constrained Multi-Objective Markov Decision Process (MOMDP) that is coupled to, but does not replace, the base DRL task policy. The base agent observes the environment state $s_t$ and selects the environment action $a^{env}_t$ using its own task policy $\pi^{task}_{\psi}$. \TetraRL{} observes hardware telemetry, resource state, and the current preference vector, and selects only the runtime action $a^{sys}_t$. Thus, the runtime state is
\begin{equation}
z_t = [h_t,\; q_t,\; f^{cpu}_t,\; f^{gpu}_t,\; \omega_t],
\end{equation}
where $h_t$ contains latency, memory, power/energy, and temperature telemetry; $q_t$ contains algorithm-specific queue or buffer state; and $\omega_t\in\Delta^3$ is the four-objective preference vector. The runtime action is
\begin{equation}
a^{sys}_t = (\Delta u_t,\; \Delta f^{cpu}_t,\; \Delta f^{gpu}_t),
\end{equation}
where $\Delta u_t$ changes the algorithm-specific resource knob, such as replay capacity, minibatch size, rollout length, or update ratio. The environment action $a^{env}_t$ is not selected by \TetraRL{}.

\subsection{4-D Reward Vector and Scalarization}

The instantaneous reward is a 4-D vector with deliberately heterogeneous units:

\begin{equation}
\mathbf{r}_t =
\big[-c_L(L_t),\; \bar r_{env,t},\; -c_M(M_t),\; -c_E(E_t)\big],
\end{equation}

Because the four components live on different scales, we apply per-dimension running z-score normalization with a 1{,}000-step warmup window and an exponentially-weighted moving-average filter on the raw telemetry, mirroring the empirical recommendations in multi-objective reinforcement learning (MORL) work, i.e., PD-MORL~\cite{basaklar2023pdmorl} and Spoor et al.~\cite{spoor2025lagrangian}. For a given preference $\omega$, the scalar surrogate is $r_{scalar,t} = \omega^\top \mathbf{r}_t$. Following the C-MORL~\cite{liu2025cmorl} two-stage philosophy, the constraint terms can also be re-expressed as inequalities $\mathbb{E}[c_E] \le \tau_E$, $\mathbb{E}[c_M] \le \tau_M$ to be handled by Lagrangian dual variables; the Override Layer  acts as a final hardware-enforced safety net below the algorithmic Lagrangian.

\subsection{Hypervolume as the Primary Metric}

We follow the MORL community~\cite{hayes2022morlsurvey,vamplew2011empirical,zitzler2007hv} in adopting the hypervolume (HV) indicator $\mathcal{H}(F;r_*)$ of the achieved Pareto front $F$ relative to a reference point $r_*$. HV is the dominated volume of the rectangle $[r_*,p]$ for every $p \in F$, and is monotone in Pareto dominance: a strictly better front always has strictly higher HV. For the Deep Sea Treasure (DST) 2-D benchmark in our Section~\ref{sec:eval} experiments we use the standard reference point $r_* = (0.0, -25.0)$. Concretely, for a 2-D non-dominated front $F = \{(R^{(i)}_1, R^{(i)}_2)\}_{i=1}^{|F|}$ sorted by $R_1$ ascending, we compute
\begin{equation}
\mathcal{H}(F; r_*) = \sum_{i=1}^{|F|} \big(R^{(i)}_1 - R^{(i-1)}_1\big)\,\big(R^{(i)}_2 - r_{*,2}\big)\,\mathbf{1}\!\left[R^{(i)}_2 > r_{*,2}\right],
\label{eq:hv}
\end{equation}
with the convention $R^{(0)}_1 = r_{*,1}$. We log HV jointly with the front cardinality $|F|$ and the sparsity (mean inter-point Euclidean distance), so that a high HV achieved by a degenerate single-point front is detected and reported as such.

\subsection{Soft-Constraint Lagrangian and Override Composition}

Within a fixed preference $\omega$, the constrained MOMDP induces the per-episode optimization
\begin{equation}
\begin{aligned}
\max_{\pi} \;\; & \mathbb{E}_{\pi}\!\left[\sum_{t} \gamma^t \omega^\top \mathbf{r}_t\right]\\
\text{s.t.}\;\; & \mathbb{E}_{\pi}\!\left[\sum_{t} \gamma^t c_k(s_t,a_t)\right] \le \tau_k,\\
                & \forall k\in\{E,M\}.
\end{aligned}
\label{eq:constrained}
\end{equation}
Eq.~(\ref{eq:constrained}) can be solved by primal-dual updates with adaptive Lagrangian multipliers $\lambda_k$, mirroring PPO-Lagrangian as wrapped by OmniSafe~\cite{ji2023omnisafe}. The Override Layer is composed \emph{below} this dual update: it does not change $\lambda_k$ but vetoes the actions the policy emits whenever telemetry crosses the hardware safety threshold, so violations of $\tau_k$ are upper-bounded by the Override's own threshold $\tau_k^{HW}$ rather than by whatever value the dual learns. 
In practice, because both DVFS transitions and on-device telemetry are subject to non-negligible actuation and sensing latency, the Override Layer cannot enforce $\tau_k^{HW}$ instantaneously. We therefore position it as an \emph{empirical recovery layer} that substantially reduces the frequency and magnitude of constraint violations rather than as a hard, zero-violation guarantee; the residual violations reported in our evaluation tables are a direct consequence of this delayed actuation, and the bound above should be read as the target that the Override drives the system back toward once a threshold crossing is detected.

\section{The \TetraRL Framework Architecture}
\label{sec:framework}
\begin{figure}[!htbp]
\centering
\includegraphics[width=0.92\textwidth]{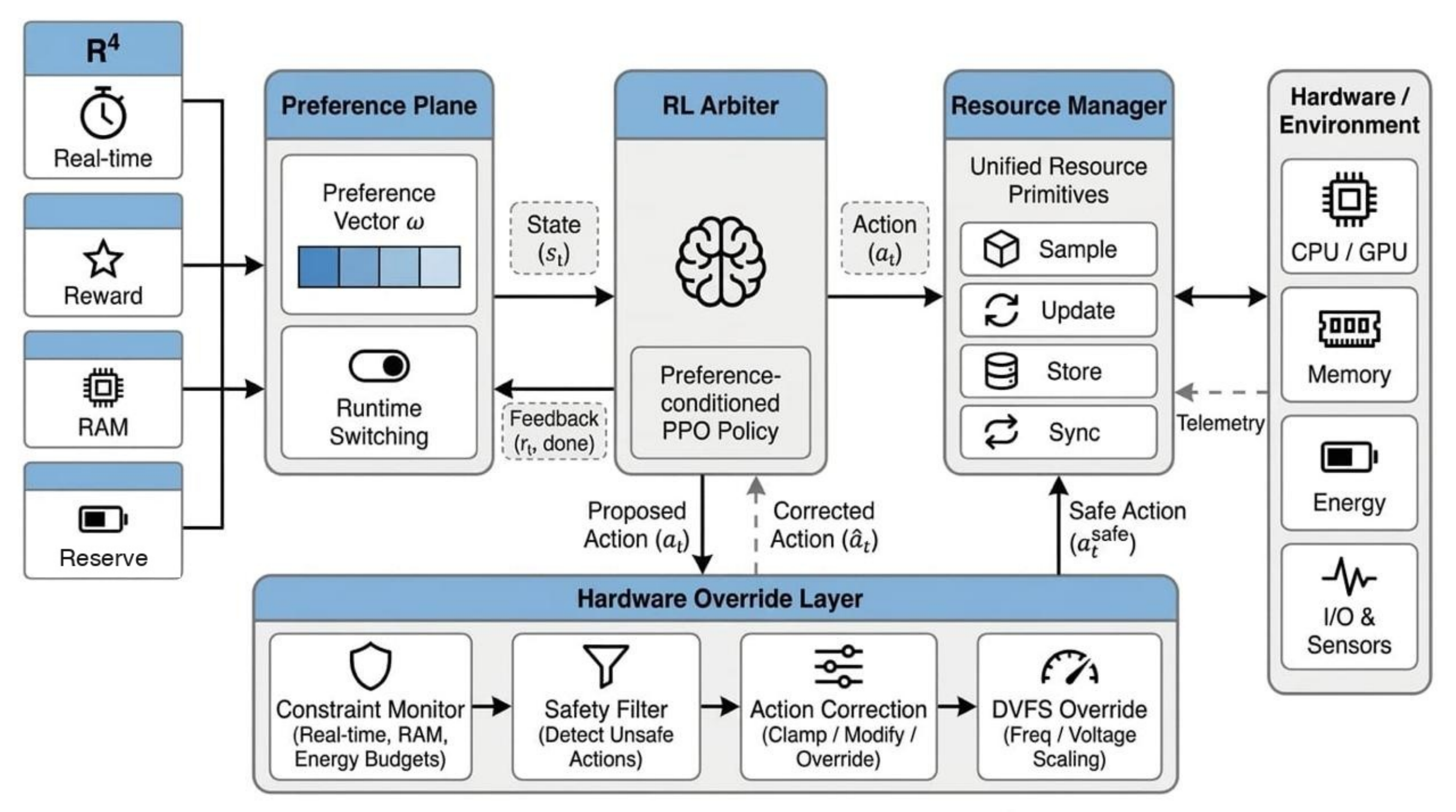} 

\caption{\TetraRL{} runtime architecture. The base DRL agent chooses environment actions, while \TetraRL{} controls runtime actions. The Preference Plane emits $\omega_t$; the Resource Manager builds runtime state $z_t$ from telemetry and current resource configuration; the RL Arbiter proposes a system action $a^{sys}_t$; and the Hardware Override Layer projects unsafe proposals to an executed action $\tilde{a}^{sys}_t$. The Resource Manager maps $\tilde{a}^{sys}_t$ onto unified resource primitives: Sample $C_{\text{sample}}$, Update $C_{\text{update}}$, Store $M_{\text{store}}$, and Sync $\tau_{\text{sync}}$, and applies the corresponding runtime and DVFS updates on Jetson hardware. Telemetry closes the loop by updating the next runtime state.}
\label{fig:arch}
\end{figure}

Section~\ref{sec:problem} separates the base DRL task policy from the \TetraRL{} runtime policy. In this section, we describe how this runtime controller is implemented. At each control point, the Resource Manager collects telemetry and constructs the runtime state $z_t$; the Preference Plane provides the current objective weights $\omega_t$; the RL Arbiter proposes a system action $a^{sys}_t$; and the Hardware Override Layer projects unsafe proposals to an executed action $\tilde{a}^{sys}_t$. The Resource Manager then applies $\tilde{a}^{sys}_t$ through algorithm-specific resource knobs and CPU-GPU DVFS controls, while the base DRL agent continues to choose environment actions through its own task policy. Fig.~\ref{fig:arch} demonstrates \TetraRL{} architecture.

\subsection{Four-Component Decomposition}

\TetraRL{} decomposes runtime control into four components with explicit interfaces:

\begin{enumerate}[leftmargin=14pt]
\item \textbf{Preference Plane.} Produces the preference vector $\omega_t$ in the \Rfour{}: real-time, reward, memory, and energy. During training, $\omega_t$ is sampled from $\mathrm{Dirichlet}(\mathbf{1})$ at episode boundaries to expose the Arbiter to different operating regimes. During deployment, $\omega_t$ may be supplied by an application planner or scheduler, e.g., shifting toward energy under low battery or toward latency under a tighter control deadline.

\item \textbf{Resource Manager.} Collects runtime telemetry, including latency, memory, power, current resource configuration, and CPU, GPU DVFS levels. It forms the runtime state $z_t$ used by the Arbiter and applies the executed system action $\tilde{a}^{sys}_t$ to algorithm-specific resource knobs and DVFS controls. DVFS changes are applied at block granularity to amortize transition latency.

\item \textbf{RL Arbiter.} Implements the preference-conditioned runtime policy $\pi^{sys}_{\theta}$. The Arbiter observes $z_t$ and proposes a system action $a^{sys}_t$, such as changing the algorithm-specific resource knob or moving CPU/GPU frequencies through the available DVFS tables. The base DRL agent separately selects the environment action.

\item \textbf{Hardware Override Layer.} Acts as a conservative runtime guardrail. Given $a^{sys}_t$ and $z_t$, it checks deployment thresholds and replaces unsafe proposals with a recovery action $\tilde{a}^{sys}_t$ when necessary. Because telemetry and actuation are delayed, the Override is evaluated as a mechanism for reducing violation frequency and magnitude rather than as a proof of zero violations.
\end{enumerate}

\subsection{Unified Resource-Primitive Abstraction}

A central design choice in \TetraRL{} is to expose different DRL algorithms through a common Resource Manager interface. The interface consists of four resource primitives: $C_{\text{sample}}$ for environment interaction, $C_{\text{update}}$ for gradient-update compute, $M_{\text{store}}$ for replay or rollout storage, and $\tau_{\text{sync}}$ for synchronization overhead. Each DRL algorithm implements a lightweight wrapper that maps its native knobs onto these primitives. For off-policy agents, these knobs may include replay minibatch size, replay capacity, and update ratio; for on-policy agents, they may include rollout length, minibatch size, and update schedule. Table~\ref{tab:primitives} maps the evaluated DRL algorithms onto the four primitives. The Resource Manager interface is shared across algorithms, while each algorithm supplies a lightweight wrapper that maps its native knobs to the primitive interface.

\begin{table}[!htbp]
\centering
\caption{Mapping of evaluated DRL algorithms onto \TetraRL{}'s four resource primitives. Off-policy methods expose replay and target-network costs, while on-policy methods expose rollout, minibatch, and synchronization costs.}
\label{tab:primitives}
\renewcommand{\arraystretch}{1.15}
\scriptsize
\setlength{\tabcolsep}{3pt}
\resizebox{0.8\columnwidth}{!}{
\begin{tabular}{l l l l l}
\toprule
Alg. & $C_{\text{sample}}$ & $C_{\text{update}}$ & $M_{\text{store}}$ & $\tau_{\text{sync}}$\\
\midrule
DQN  & env.step + replay  & Q-network SGD        & replay buf.  & target-net sync\\
DDQN & env.step + replay  & double-Q SGD         & replay buf.  & target-net sync\\
C51  & env.step + replay  & distributional SGD   & replay buf.  & target-net sync\\
A2C  & rollout collection & actor-critic update  & rollout buf. & \texttt{n\_envs} sync\\
PPO  & rollout collection & clipped policy update & rollout buf. & \texttt{n\_envs} sync\\
\bottomrule
\end{tabular}
}
\end{table}

\subsection{Preference-Conditioned Runtime Policy}

The RL Arbiter implements the runtime policy $\pi^{sys}_{\theta}$ from Section~\ref{sec:problem}. Following preference-conditioned MORL methods such as PCN~\cite{reymond2022pcn} and PD-MORL~\cite{basaklar2023pdmorl}, the actor and critic condition on the runtime state, which already includes the preference vector:
\begin{equation}
\pi^{sys}_{\theta}(a^{sys}\mid z_t),\qquad
V_{\phi}(z_t),
\label{eq:policy}
\end{equation}
where $z_t=[h_t,q_t,f^{cpu}_t,f^{gpu}_t,\omega_t]$. The actor outputs changes to the algorithm-specific resource knob and DVFS levels, while the base DRL agent separately selects the environment action. PPO updates use the scalarized runtime reward. During training, $\omega_t$ is sampled from $\mathrm{Dirichlet}(\mathbf{1})$ at episode boundaries and held fixed within the episode; during evaluation, we sweep fixed preference anchors to recover the achieved Pareto front. Because the two policies act on disjoint action spaces and share no parameters, \TetraRL{} is \emph{transparent} to the base DRL agent: it leaves the agent's observation, network, loss, and environment-action selection untouched and only reshapes how much compute and energy the surrounding training loop may consume, so disabling the Arbiter recovers the original training procedure unchanged. 

At evaluation time, the Arbiter is frozen and evaluated over a fixed sweep of preference anchors covering the simplex corners and interior. For each anchor, the policy runs deterministically for $m$ episodes, producing normalized return vectors in the canonical \Rfour{} order. We then log HV, front cardinality, and sparsity when available, and save the best checkpoint according to the front-level metric.

\begin{algorithm}[!t]
\caption{\TetraRL{} Runtime-Control Loop}
\label{alg:tetrarl}
\begin{algorithmic}[1]
\State Init base DRL agent $\pi^{task}_{\psi}$, Arbiter $\pi^{sys}_{\theta}$, Resource Manager $RM$, Override $OL$
\For{episode $e=1..E$}
  \State Sample or receive preference $\omega_e$
  \For{control block $b=1..B$}
    \State $h_t \gets RM.\textsc{Telemetry}()$
    \State $z_t \gets [h_t,q_t,f^{cpu}_t,f^{gpu}_t,\omega_e]$
    \State $a^{sys}_t \sim \pi^{sys}_{\theta}(\cdot \mid z_t)$
    \State $\tilde{a}^{sys}_t \gets OL.\textsc{Project}(a^{sys}_t,z_t)$
    \State $RM.\textsc{Apply}(\tilde{a}^{sys}_t)$
    \For{step $j=1..N_{block}$}
      \State $a^{env}_j \sim \pi^{task}_{\psi}(\cdot \mid s_j)$
      \State $s_{j+1}, r^{env}_j \gets \mathrm{env.step}(a^{env}_j)$
      \State Update base DRL agent w.r.t. its algorithm
    \EndFor
    \State Measure $L_t,M_t,E_t$ and compute $r^{sys}_t$
    \State Store $(z_t,\tilde{a}^{sys}_t,r^{sys}_t,z_{t+1})$ for Arbiter update
  \EndFor
  \State Update $\pi^{sys}_{\theta}$ with PPO using scalarized runtime rewards
  \If{evaluation point}
    \State Run preference-anchor sweep and log HV, $|F|$, and sparsity
  \EndIf
\EndFor
\end{algorithmic}
\end{algorithm}

\subsection{Action Masking for Real-Time Feasibility}

A naive Arbiter can propose DVFS settings and resource-knob combinations that are predicted to violate latency, memory, or energy thresholds under the current runtime state. Before sampling, \TetraRL{} constructs a conservative feasible action set $\mathcal{A}_{feas}(z_t)\subseteq\mathcal{A}^{sys}$ using current telemetry, reserved memory headroom, and profiled DVFS latency/energy estimates. The Arbiter assigns $-\infty$ logits to actions outside this set before the softmax, following standard invalid-action masking practice~\cite{schulman2017ppo}. This mask reduces obviously unsafe proposals, while the Override Layer remains responsible for correcting residual unsafe actions caused by telemetry delay, workload drift, or profiling error. Concretely, an action $(\Delta b, \Delta f_{cpu}, \Delta f_{gpu})$ is masked if either: 
\begin{itemize}[leftmargin=12pt]
\item the projected per-step latency under the proposed frequencies (read from the offline DVFS profile in Table~\ref{tab:dvfs_orin}) exceeds the remaining per-step latency budget (this assumes the workload is well-modeled by the offline profile, which is a stronger assumption for DRL training, where the workload drifts as the policy converges, than for inference; or
\item the projected unified-memory footprint under the new control knob exceeds $M_{max} - M_{reserved}$.
\end{itemize} 

\subsection{Hardware Override Layer}
\label{sec:masking}

The Override Layer is a practical, empirical integration in our framework rather than a formally guaranteed safety filter: because telemetry sensing and DVFS actuation are both delayed, it cannot promise zero constraint violations, and we present it only as a mechanism that reduces the frequency and magnitude of violations in practice. It is placed after the Arbiter and before action execution. Given a proposed system action $a^{sys}_t$ and runtime state $z_t$, it checks latency, memory, and energy telemetry against deployment thresholds with safety margins. If the proposal is predicted to be unsafe, the Override replaces it with a conservative recovery action selected from the profiled feasible set:

\[
\tilde{a}^{sys}_t =
\begin{cases}
a^{sys}_t, & \text{if } a^{sys}_t \in \mathcal{A}_{feas}(z_t),\\
a^{rec}_t, & \text{otherwise.}
\end{cases}
\]
The recovery action is chosen to reduce the active violation while preserving latency feasibility when possible. If no feasible recovery action exists under the current profile, the run is marked constraint-infeasible. Consider one concrete energy-driven recovery example. Suppose the Arbiter is at $(b{=}256, f_{cpu}{=}2.20\,\mathrm{GHz},$ $ f_{gpu}{=}1.30\,\mathrm{GHz})$ and proposes $a_t=(\Delta b{=}+0, \Delta f_{cpu}{=}+0, \Delta f_{gpu}{=}+0)$, but the Resource Manager's freshest EMA reports $E_{step,t} = 5.4\,\mathrm{J} > E_{max} = 5.0\,\mathrm{J}$. The Override emits $a_{safe} = (b{=}64, f_{cpu}{=}115.2\,\mathrm{MHz},$ $ f_{gpu}{=}306\,\mathrm{MHz})$. The lower GPU frequency \emph{does} raise the per-step latency (a deliberately conservative trade), but the smaller batch size shortens the GPU work-per-step enough that the next-step latency budget still has positive slack at the offline-profile latency for $(64, 115.2, 306)$; the policy then re-enters the loop with a lower energy reading and, if the EMA cools below $E_{max}$, the Arbiter is free to climb back up. The fixed point of the recovery is the operating point at which $E_{step}$ falls below $E_{max}$ and the projected latency at the recovered $(b,f_{cpu},f_{gpu})$ still meets the per-step latency budget; if no such point exists in the offline profile the Override remains pinned at the lowest corner and the run is reported as a constraint-infeasible run.

\subsection{Algorithm Details}

Algorithm~\ref{alg:tetrarl} shows the full training loop in detail. We walk through it line by line. \emph{Line~1} initializes the four interacting components: the base task agent $\pi^{task}_{\psi}$ that learns the control policy for the environment, the system Arbiter $\pi^{sys}_{\theta}$ that decides runtime control knobs, the Resource Manager $RM$ that reads telemetry and actuates hardware, and the Override Layer $OL$ that filters unsafe proposals. \emph{Line~2} opens the outer loop over training episodes. \emph{Line~3} samples or receives the preference vector $\omega_e$ that scalarizes the multi-objective trade-off (return versus latency, memory, and energy) for the episode, so the same network can be conditioned on different operating points. \emph{Line~4} opens the inner loop over control blocks: the Arbiter acts at the coarser block granularity rather than every environment step, which keeps DVFS and knob changes infrequent enough to avoid actuation thrash.

Within each control block, \emph{line~5} queries the Resource Manager for the current hardware telemetry $h_t$, and \emph{line~6} assembles the Arbiter's observation $z_t$ by concatenating that telemetry with the queue/state features $q_t$, the current CPU and GPU frequencies $f^{cpu}_t,f^{gpu}_t$, and the active preference $\omega_e$. \emph{Line~7} samples a system action $a^{sys}_t$ (a DVFS/knob adjustment) from the Arbiter conditioned on $z_t$. \emph{Line~8} passes that proposal through the Override Layer's \textsc{Project} operation, which masks or replaces it with a conservative feasible action $\tilde{a}^{sys}_t$ when the raw proposal is predicted to violate a constraint; \emph{line~9} then applies the projected, vetted action to the hardware through the Resource Manager.

\emph{Lines~10--14} run the base DRL agent for $N_{block}$ environment steps under the just-applied hardware configuration: \emph{line~11} samples the environment action $a^{env}_j$ from the task policy, \emph{line~12} steps the environment to obtain the next state and reward, and \emph{line~13} updates the base agent according to its own learning rule (e.g.\ the underlying DQN or PPO update), keeping the task learner and the system Arbiter cleanly separated. After the block finishes, \emph{line~15} measures the realized latency, memory, and energy $L_t,M_t,E_t$ and computes the Arbiter's scalarized system reward $r^{sys}_t$, and \emph{line~16} stores the transition $(z_t,\tilde{a}^{sys}_t,r^{sys}_t,z_{t+1})$ for the Arbiter's update. Note that the reward and transition are logged against the \emph{projected} action $\tilde{a}^{sys}_t$ actually executed, not the raw proposal, so the Arbiter learns from what was really applied.

Once all control blocks in the episode complete (\emph{line~17}), \emph{line~18} updates the Arbiter $\pi^{sys}_{\theta}$ with PPO on the collected runtime transitions using the scalarized rewards. Finally, \emph{lines~19--21} run the periodic evaluation: at an evaluation point the algorithm performs the preference-anchor sweep and logs the hypervolume HV, the front size $|F|$, and the sparsity of the resulting Pareto front, which serve as the primary multi-objective quality metrics. \emph{Line~22} closes the episode loop.

\section{Implementation}
\label{sec:impl}

\subsection{Tegrastats Async Daemon and DVFS Controller}

The tegrastats async daemon is a 100\,Hz sampler that scrapes the \texttt{tegrastats} stream, parses CPU power, GPU power, EMC bandwidth, and per-rail temperature, and dispatches a 10\,Hz EMA-filtered summary to the Resource Manager via a lock-free ring buffer. The kernel/user split (sampler runs as a low-priority daemon, dispatcher polls from the RL process) follows the DVFS-DRL-Multitask pattern~\cite{dvfsdrl2024} and bounds the worst-case sensing overhead to a few hundred microseconds per RL step.

The DVFS controller writes to the cpufreq sysfs setspeed node after first switching the governor to \texttt{userspace}; on shutdown we reset to \texttt{schedutil}. We measured the actual per-transition latency for every (from, to) frequency pair on Orin AGX (Table~\ref{tab:dvfs_orin}). The CPU table has $29 \times 28 = 812$ distinct transitions averaging 0.243\,ms (max 0.374\,ms); the GPU table has $11 \times 10 = 110$ transitions averaging 3.399\,ms (max 6.857\,ms). The asymmetry (GPU is roughly $14\times$ slower than CPU per transition) is what motivates the super-block decision granularity: one DVFS adjustment per $N{=}10$ training steps amortizes a 6.857\,ms worst-case GPU transition over a 50--500\,ms training step block.

\subsection{Pre-Allocated Soft-Truncation Replay Buffer}

PyTorch on Orin's unified memory fragments quickly under repeated allocate/free of large tensors~\cite{li2023r3}. The replay buffer is pre-allocated at maximum size at start-up; runtime ``soft truncation'' is implemented by an index mask that tells the sampler to draw only from a prefix of the physically-allocated array. This eliminates the fragmentation pathology of repeated allocate/free cycles on unified memory while preserving the algorithmic effect of resizing.

\subsection{Codebase and Reproducibility}

\TetraRL is implemented in $\sim$5\,KLOC of Python on top of PyTorch~\cite{paszke2019pytorch}, Gymnasium~\cite{gymnasium2023}, and the multi-objective extension MO-Gymnasium~\cite{mogymnasium2023}, with a single-process \texttt{cleanrl}~\cite{huang2022cleanrl} PPO backbone driving the preference-conditioned arbiter. The on-device head-to-head evaluation spans three benchmarks: \texttt{CartPole-v0} and Atari-Breakout inherited from R$^3$~\cite{li2023r3} plus the higher-variance \texttt{CartPole-v1}, each crossed with five DRL algorithms (DQN, DDQN, C51, A2C, PPO) and five runtime wrappers (MAX-A, MAX-P, R$^3$, DuoJoule, and \TetraRL{}), executed on two NVIDIA Jetson platforms (Orin AGX and Orin Nano) running the JetPack~6.2 CUDA stack with PyTorch~2.8. To make every reported number reproducible, unless stated otherwise, each cell fixes its compute budget rather than its episode count ($5\times10^{4}$ samples for CartPole-v0, $3.2\times10^{6}$ for CartPole-v1, and $1.28\times10^{7}$ for Breakout); CartPole cells report mean$\pm$std over three seeds while the heavier Breakout sweeps follow a single-seed protocol, and all latency, energy, and peak-memory figures are collected from on-board telemetry. The complete experimental harness, runner scripts, per-cell launch commands, DVFS profiles, and the telemetry-aggregation pipeline, is staged for release on acceptance to enable end-to-end reproduction on commodity Jetson hardware.

\subsection{Hyperparameters and Training Schedule}

Unless stated otherwise, \TetraRL trains with the following defaults, chosen to match PD-MORL~\cite{basaklar2023pdmorl} and \texttt{cleanrl}~\cite{huang2022cleanrl} reference settings so that any HV deltas are attributable to the framework rather than to undisclosed tuning. Actor and critic are 3-layer MLPs of widths $(256,256,128)$; the third 128-wide layer adds a separate $\omega$-conditioning bottleneck without altering the (256, 256) trunk that PD-MORL uses, and is the only network-width deviation from PD-MORL's reference. On DAG environments, the first two MLP layers are replaced by a 64-dim GraphSAGE encoder. The PPO clip ratio is $0.2$, the entropy coefficient $0.01$, the value coefficient $0.5$, GAE $\lambda=0.95$, and the PPO horizon \texttt{n\_steps}=2048 with 10 epochs of minibatch size 64. The Adam learning rate is $3\!\times\!10^{-4}$ for the actor and $1\!\times\!10^{-3}$ for the critic. Preferences are sampled from $\mathrm{Dirichlet}(\mathbf{1})$ at every episode boundary and held fixed within an episode. Evaluation runs at $K_{eval}=10{,}000$ environment steps with $m=5$ episodes per anchor over an 11-anchor sweep. The DVFS super-block size is $N_{block}=10$, chosen so that one worst-case 6.857\,ms GPU transition (Table~\ref{tab:dvfs_orin}) is amortized over a $\sim$50\,ms training-step block (i.e.\ $<$15\% overhead).

\begin{figure}[!htbp]
\centering
\includegraphics[width=\linewidth]{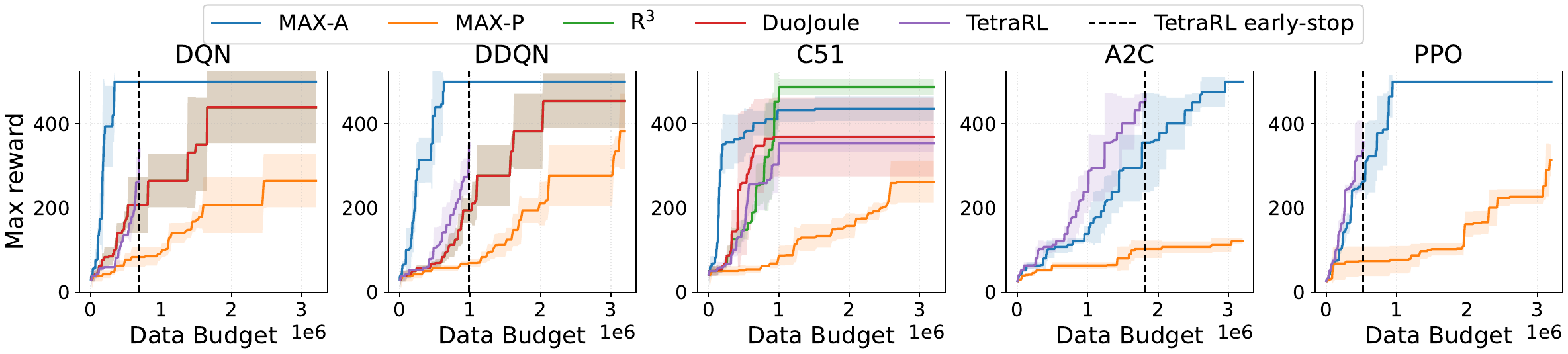}
\caption{Per-episode running-max reward vs data budget for \texttt{CartPole-v1} on Orin AGX under the controlled-variable protocol. Every cell consumes the same total data budget ($3.2\times10^{6}$ samples). Each curve is the running maximum of per-episode reward (best-so-far); per-episode reward is clipped to the environmental upper bound of 500. The black vertical dashed line marks \TetraRL{}'s reward-ceiling early-stop position. Note that R$^3$ and DuoJoule do not apply to A2C/PPO cells.}
\label{fig:p15_phase1_reward}
\end{figure}

\subsection{Reward-Ceiling Early-Exit Hook}
\label{sec:early_exit_impl} 

Most on-device DRL baselines~\cite{li2023r3,duojoule2024} treat the training horizon as a fixed sample budget: they continue to step the environment and apply gradient updates until the budget is exhausted, even after the policy has already attained the environment's reward ceiling. On capped-return tasks this wastes both wall-clock latency and board-input energy on episodes that cannot improve the reward column further, and in value-based agents it actively risks catastrophic forgetting because the replay distribution keeps shifting after the task is solved. Our implementation of \TetraRL, therefore adds a lightweight \emph{reward-ceiling early-exit} hook as part of the Arbiter's stopping logic. For environments with a known per-episode reward upper bound $R_{\max}$ (e.g.\ CartPole-v0 with $R_{\max}{=}200$, CartPole-v1 with $R_{\max}{=}500$), the Arbiter monitors the cumulative reward of each completed episode and triggers an exit as soon as a single episode return satisfies $R_t \geq R_{\max}$:

\begin{equation}
\text{stop training at step } s \iff \exists\, t \leq s :\; R_t \geq R_{\max}.
\label{eq:early_exit}
\end{equation}

This single-episode trigger is intentional: once the policy has demonstrated it can solve the task once, additional training cannot raise the reward column but can only consume more samples, energy, and wall time. The Arbiter writes an early exit flag, the triggering episode index, and the step count into the run summary, so downstream telemetry (latency, energy, peak RAM) reflects the actual work performed rather than the nominal budget. For environments without a known bounded ceiling, the hook is a no-op and \TetraRL falls back to running to the full budget like every baseline.

\subsection{What We Inherited from \Rthree and DuoJoule}

Two pieces of \Rthree~\cite{li2023r3} are reused verbatim. The first is the EMA-filtered moving-average controller for the data-tracking metric, which we generalize from a single $D_n$ data-rate target to an EMA over the entire $h(t)$ vector. The second is the FFmpeg co-runner protocol at 720p / 1080p / 2K, which we re-use as the OS-interference reference workload (the FFmpeg-corunner cross-platform CDF specifically is still deferred; per-step framework latency is reported by hardware tier in Section~\ref{sec:eval}). From DuoJoule~\cite{duojoule2024} we reuse the kernel-userspace split for low-overhead sensing and the per-DNN MetricTracker pattern (renamed Resource Manager in our codebase). We did \emph{not} reuse DuoJoule's heuristic ``efficiency score'' or its discrete energy-mode switcher: both are subsumed by the preference-conditioned Arbiter, since any efficiency-score level is recoverable as a specific $\omega$ in our simplex.

\section{Evaluation}
\label{sec:eval}

We evaluate three claims: (i) a single \TetraRL policy can recover a high-quality Pareto front on a standard MORL benchmark, (ii) the front transfers to Orin AGX hardware with low evaluation latency, and (iii) the four-component framework's overhead is small enough for embedded deployment.

\subsection{Experimental Setup}
\label{sec:experimental_setup}

\noindent\textbf{Setup overview.} We evaluate \TetraRL{} on two NVIDIA Jetson platforms (AGX Orin 32\,GB, Orin Nano 8\,GB) across four DRL environments: CartPole-v0, CartPole-v1, Atari Breakout NoFrameskip-v4, and a synthetic DAG-scheduler multi-objective  environment), five DRL algorithms (DQN, DDQN, C51, A2C, PPO), and five training-runtime wrappers (MAX-A, MAX-P, R$^3$, DuoJoule, \TetraRL{}).

\noindent\textbf{Hardware.} We adopt an NVIDIA Jetson AGX Orin and an NVIDIA Jetson Orin Nano as a testbed. On both platforms the CPU governor is pinned to \texttt{userspace} during DVFS-active runs and reset to \texttt{schedutil} on shutdown. 

\noindent\textbf{Software stack.} \TetraRL{} is implemented in $\sim$5\,KLOC of Python on top of PyTorch~\cite{paszke2019pytorch}, Gymnasium~\cite{gymnasium2023}, and the multi-objective extension MO-Gymnasium~\cite{mogymnasium2023}, with a single-process \texttt{cleanrl}~\cite{huang2022cleanrl} PPO backbone for the preference-conditioned arbiter. The CartPole, and dag\_scheduler\_mo benchmarks are based on the \texttt{mo-gymnasium} suite; Atari Breakout is wrapped through the ALE-py preprocessing stack inherited from R$^3$~\cite{li2023r3}. Both Jetson platforms run the JetPack~6.x CUDA stack with PyTorch~2.8. The complete experimental harness toolchain is staged for release on acceptance.

\noindent\textbf{Environments.} Following R$^3$ and DuoJoule, we leverage CartPole-v0 and Atari Breakout as simple DRL environments for standard benchmarking. We extend a more modern and stability-sensitive CartPole-v1 environment in our testing. Besides, we leverage \texttt{dag\_scheduler\_mo}, a standard MORL benchmark from \texttt{mo-gymnasium} for verifying Pareto effectiveness.

\noindent\textbf{DRL Algorithms.} Five DRL algorithms are evaluated across the on-device head-to-head: \emph{DQN}~\cite{mnih2015dqn} and \emph{DDQN}~\cite{van2016deep} (off-policy value-based, replay-buffer driven); \emph{C51}~\cite{bellemare2017distributional} (distributional value-based, 51-atom categorical head); \emph{A2C}~\cite{mnih2016asynchronous} and \emph{PPO}~\cite{schulman2017ppo} (on-policy actor-critic, rollout-buffer driven). The preference-conditioned arbiter on top of the lightweight \texttt{cleanrl} backbone uses PPO.

\begin{table}[!htbp]
\centering
\caption{CartPole-v0 experimental results on Orin AGX. Compute budget held constant at $5\times10^{4}$ samples per cell, following the setting in R$^3$~\cite{li2023r3}. All metrics are end-to-end aggregated over 3 seeds as mean $\pm$ std. Bold marks the per-algo best mean per column; underline marks the per-algo second-best; ties within 0.5\% are co-marked.}
\label{tab:p15_phase1_cartpole_v0}
\renewcommand{\arraystretch}{1.05}
\scriptsize
\setlength{\tabcolsep}{3pt}
\resizebox{0.7\columnwidth}{!}{
\begin{tabular}{llrrrr}
\toprule
Algo & Wrapper & Latency (s) & Energy (J) & Peak RAM (MB) & Max reward \\
\midrule
DQN & MAX-A & \underline{193.1 $\pm$ 0.7} & \underline{130.2 $\pm$ 0.6} & \underline{2909 $\pm$ 3} & \textbf{200.00 $\pm$ 0.00} \\
DQN & MAX-P & 197.3 $\pm$ 4.1 & 133.4 $\pm$ 2.7 & 2944 $\pm$ 45 & \textbf{200.00 $\pm$ 0.00} \\
DQN & R$^3$ & 200.5 $\pm$ 3.9 & 135.9 $\pm$ 1.9 & \underline{2898 $\pm$ 7} & \textbf{200.00 $\pm$ 0.00} \\
DQN & DuoJoule & 199.8 $\pm$ 2.5 & 136.0 $\pm$ 1.6 & \underline{2901 $\pm$ 5} & \textbf{200.00 $\pm$ 0.00} \\
DQN & \textbf{TetraRL} & \textbf{19.7 $\pm$ 12.3} & \textbf{9.3 $\pm$ 8.3} & \textbf{2882 $\pm$ 3} & \textbf{200.00 $\pm$ 0.00} \\
\midrule
DDQN & MAX-A & \underline{211.2 $\pm$ 2.9} & \underline{132.0 $\pm$ 2.3} & 2993 $\pm$ 11 & \textbf{200.00 $\pm$ 0.00} \\
DDQN & MAX-P & 215.2 $\pm$ 3.9 & 133.5 $\pm$ 1.7 & 2925 $\pm$ 52 & \textbf{200.00 $\pm$ 0.00} \\
DDQN & R$^3$ & 217.7 $\pm$ 0.3 & 135.9 $\pm$ 0.6 & \underline{2894 $\pm$ 3} & \textbf{200.00 $\pm$ 0.00} \\
DDQN & DuoJoule & 218.4 $\pm$ 2.9 & 135.4 $\pm$ 1.3 & 2925 $\pm$ 47 & \textbf{200.00 $\pm$ 0.00} \\
DDQN & \textbf{TetraRL} & \textbf{13.7 $\pm$ 2.0} & \textbf{4.5 $\pm$ 0.9} & \textbf{2870 $\pm$ 6} & \textbf{200.00 $\pm$ 0.00} \\
\midrule
C51 & MAX-A & \underline{319.9 $\pm$ 5.1} & \underline{166.9 $\pm$ 2.7} & \underline{3086 $\pm$ 3} & \textbf{200.00 $\pm$ 0.00} \\
C51 & MAX-P & 323.4 $\pm$ 4.7 & 171.0 $\pm$ 2.2 & 3154 $\pm$ 118 & \textbf{200.00 $\pm$ 0.00} \\
C51 & R$^3$ & 327.1 $\pm$ 2.4 & 172.6 $\pm$ 3.0 & 3140 $\pm$ 81 & \textbf{200.00 $\pm$ 0.00} \\
C51 & DuoJoule & 324.4 $\pm$ 4.2 & 171.3 $\pm$ 2.2 & 3137 $\pm$ 106 & \textbf{200.00 $\pm$ 0.00} \\
C51 & \textbf{TetraRL} & \textbf{42.2 $\pm$ 31.8} & \textbf{16.7 $\pm$ 16.6} & \textbf{3057 $\pm$ 11} & \textbf{200.00 $\pm$ 0.00} \\
\midrule
A2C & MAX-A & 121.9 $\pm$ 2.1 & 521.5 $\pm$ 7.5 & \textbf{3142 $\pm$ 8} & \textbf{200.00 $\pm$ 0.00} \\
A2C & MAX-P & \underline{118.5 $\pm$ 1.4} & \underline{507.5 $\pm$ 5.7} & \underline{3172 $\pm$ 27} & \textbf{200.00 $\pm$ 0.00} \\
A2C & R$^3$ & \multicolumn{4}{c}{\emph{N/A (incompatible)}} \\
A2C & DuoJoule & \multicolumn{4}{c}{\emph{N/A (incompatible)}} \\
A2C & \textbf{TetraRL} & \textbf{71.9 $\pm$ 34.9} & \textbf{304.6 $\pm$ 151.2} & \textbf{3141 $\pm$ 6} & \textbf{200.00 $\pm$ 0.00} \\
\midrule
PPO & MAX-A & 193.4 $\pm$ 4.9 & 519.6 $\pm$ 15.6 & \underline{3212 $\pm$ 8} & \textbf{200.00 $\pm$ 0.00} \\
PPO & MAX-P & \underline{150.7 $\pm$ 1.5} & \underline{509.6 $\pm$ 5.4} & \underline{3206 $\pm$ 2} & \textbf{200.00 $\pm$ 0.00} \\
PPO & R$^3$ & \multicolumn{4}{c}{\emph{N/A (incompatible)}} \\
PPO & DuoJoule & \multicolumn{4}{c}{\emph{N/A (incompatible)}} \\
PPO & \textbf{TetraRL} & \textbf{29.5 $\pm$ 1.3} & \textbf{76.1 $\pm$ 3.5} & \textbf{3189 $\pm$ 9} & \textbf{200.00 $\pm$ 0.00} \\
\bottomrule
\end{tabular}}
\end{table}

\begin{table}[!htbp]
\centering
\caption{CartPole-v1 experimental results on Orin AGX. Compute budget held constant at $3.2\times10^{6}$ samples per cell. All metrics are end-to-end aggregated over 3 seeds as mean $\pm$ std. Bold marks the per-algo best mean per column (lower-is-better for latency/energy/RAM; higher-is-better for reward); underline marks the per-algo second-best; ties within 0.5\% are co-marked.}
\label{tab:p15_phase1}
\renewcommand{\arraystretch}{1.05}
\scriptsize
\setlength{\tabcolsep}{3pt}
\resizebox{0.7\textwidth}{!}{
\begin{tabular}{llrrrr}
\toprule
Algo & Wrapper & Latency (s) & Energy (J) & Peak RAM (MB) & Max reward \\
\midrule
DQN & MAX-A & 1053.4 $\pm$ 5.5 & 906.5 $\pm$ 0.8 & \underline{9568 $\pm$ 59} & \textbf{500.00 $\pm$ 0.00} \\
DQN & MAX-P & \textbf{132.7 $\pm$ 1.0} & \textbf{105.2 $\pm$ 0.8} & \textbf{9495 $\pm$ 23} & \textbf{500.00 $\pm$ 0.00} \\
DQN & R$^3$ & \underline{266.6 $\pm$ 4.4} & 222.9 $\pm$ 1.3 & \underline{9608 $\pm$ 28} & \textbf{500.00 $\pm$ 0.00} \\
DQN & DuoJoule & \underline{265.3 $\pm$ 0.7} & 222.3 $\pm$ 1.0 & \textbf{9478 $\pm$ 1} & \textbf{500.00 $\pm$ 0.00} \\
DQN & \textbf{TetraRL} & 267.9 $\pm$ 1.6 & \underline{220.1 $\pm$ 1.5} & \textbf{9484 $\pm$ 13} & \textbf{500.00 $\pm$ 0.00} \\
\midrule
DDQN & MAX-A & 1129.7 $\pm$ 7.7 & 912.7 $\pm$ 5.0 & 9650 $\pm$ 21 & \textbf{500.00 $\pm$ 0.00} \\
DDQN & MAX-P & \textbf{143.7 $\pm$ 0.9} & \textbf{105.4 $\pm$ 0.7} & \textbf{9499 $\pm$ 10} & \underline{427.67 $\pm$ 125.29} \\
DDQN & R$^3$ & \underline{286.8 $\pm$ 2.4} & 223.6 $\pm$ 1.9 & \underline{9579 $\pm$ 79} & \textbf{500.00 $\pm$ 0.00} \\
DDQN & DuoJoule & 289.6 $\pm$ 1.4 & 226.0 $\pm$ 0.9 & \textbf{9485 $\pm$ 3} & \textbf{500.00 $\pm$ 0.00} \\
DDQN & \textbf{TetraRL} & \underline{286.8 $\pm$ 0.3} & \underline{221.1 $\pm$ 0.7} & \underline{9549 $\pm$ 114} & \textbf{500.00 $\pm$ 0.00} \\
\midrule
C51 & MAX-A & 1719.6 $\pm$ 12.4 & 1,093.7 $\pm$ 64.3 & 10209 $\pm$ 144 & \textbf{459.00 $\pm$ 36.29} \\
C51 & MAX-P & \textbf{218.6 $\pm$ 1.9} & \textbf{137.1 $\pm$ 0.3} & \textbf{9799 $\pm$ 22} & 314.00 $\pm$ 75.11 \\
C51 & R$^3$ & 434.3 $\pm$ 3.9 & 282.9 $\pm$ 10.3 & \underline{9959 $\pm$ 95} & 340.00 $\pm$ 131.02 \\
C51 & DuoJoule & 445.3 $\pm$ 7.5 & 288.2 $\pm$ 10.2 & \underline{9983 $\pm$ 5} & 263.33 $\pm$ 46.00 \\
C51 & \textbf{TetraRL} & \underline{430.0 $\pm$ 2.2} & \underline{261.0 $\pm$ 7.5} & 10025 $\pm$ 34 & \underline{353.67 $\pm$ 124.20} \\
\midrule
A2C & MAX-A & \underline{712.2 $\pm$ 6.9} & \underline{3,427.8 $\pm$ 36.2} & 10213 $\pm$ 23 & \textbf{500.00 $\pm$ 0.00} \\
A2C & MAX-P & \textbf{92.1 $\pm$ 0.5} & \textbf{438.2 $\pm$ 2.2} & \textbf{10075 $\pm$ 101} & \underline{219.67 $\pm$ 70.47} \\
A2C & R$^3$ & \multicolumn{4}{c}{\emph{N/A (incompatible)}} \\
A2C & DuoJoule & \multicolumn{4}{c}{\emph{N/A (incompatible)}} \\
A2C & \textbf{TetraRL} & 727.1 $\pm$ 4.2 & 3,498.7 $\pm$ 22.9 & \underline{10142 $\pm$ 96} & \textbf{500.00 $\pm$ 0.00} \\
\midrule
PPO & MAX-A & 1042.5 $\pm$ 7.3 & 3,467.3 $\pm$ 16.1 & 9975 $\pm$ 54 & \textbf{500.00 $\pm$ 0.00} \\
PPO & MAX-P & \textbf{107.9 $\pm$ 3.4} & \textbf{426.5 $\pm$ 16.7} & \underline{9759 $\pm$ 72} & \underline{388.67 $\pm$ 82.48} \\
PPO & R$^3$ & \multicolumn{4}{c}{\emph{N/A (incompatible)}} \\
PPO & DuoJoule & \multicolumn{4}{c}{\emph{N/A (incompatible)}} \\
PPO & \textbf{TetraRL} & \underline{950.7 $\pm$ 186.1} & \underline{2,999.3 $\pm$ 889.3} & \textbf{9401 $\pm$ 588} & \textbf{500.00 $\pm$ 0.00} \\
\bottomrule
\end{tabular}}
\end{table}

\begin{figure}[!htbp]
\centering
\includegraphics[width=\linewidth]{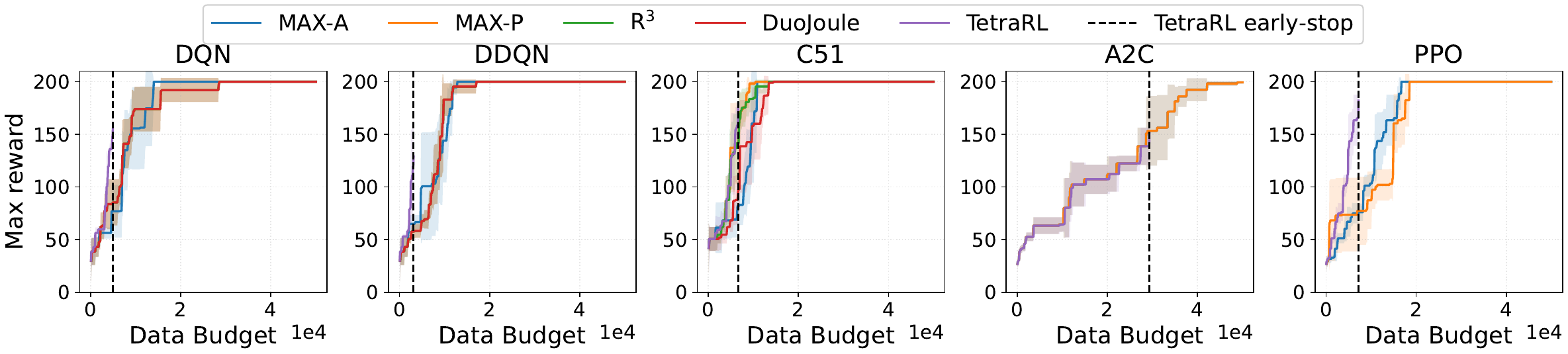}
\caption{Per-episode running-max reward vs data budget for \texttt{CartPole-v0} on Orin AGX under $5\times10^{4}$-sample setting. Each curve is the running maximum of per-episode reward (best-so-far). The black vertical dashed line marks \TetraRL{}'s reward-ceiling early-stop position. Note that R$^3$ and DuoJoule do not apply to A2C/PPO cells.}
\label{fig:p15_cartpole_v0_agx}
\end{figure}

\begin{table}[!htbp]
\centering
\caption{Atari Breakout experimental results on Orin AGX. Compute budget held constant at $1.28\times10^{7}$ samples per cell. Bold marks the per-algo best mean per column; underline marks the per-algo second-best; ties within 0.5\% are co-marked.}
\label{tab:p15_phase3}
\renewcommand{\arraystretch}{1.05}
\scriptsize
\setlength{\tabcolsep}{3pt}
\resizebox{0.7\textwidth}{!}{
\begin{tabular}{llrrrr}
\toprule
Algo & Wrapper & Latency (s) & Energy (J) & Peak RAM (MB) & Max reward \\
\midrule
DQN & MAX-A & \underline{6625.7} & 11,017.1 & 9612 & \underline{250.00} \\
DQN & MAX-P & \multicolumn{4}{c}{\emph{OOM}} \\
DQN & R$^3$ & 7029.7 & \underline{7,417.2} & \textbf{8729} & 64.00 \\
DQN & DuoJoule & 7005.7 & 7,423.5 & \textbf{8727} & 69.00 \\
DQN & \textbf{TetraRL} & \textbf{3310.2} & \textbf{5,341.1} & \underline{8911} & \textbf{320.00} \\
\midrule
DDQN & MAX-A & \underline{7271.7} & 10,956.3 & \underline{9119} & \textbf{280.00} \\
DDQN & MAX-P & \multicolumn{4}{c}{\emph{OOM}} \\
DDQN & R$^3$ & 7978.9 & \underline{7,615.4} & \textbf{8886} & 40.00 \\
DDQN & DuoJoule & 7973.9 & 7,669.6 & 9202 & 40.00 \\
DDQN & \textbf{TetraRL} & \textbf{3641.6} & \textbf{5,362.9} & 9301 & \underline{84.00} \\
\midrule
C51 & MAX-A & 9267.8 & 12,459.8 & 4923 & \textbf{32.00} \\
C51 & MAX-P & \multicolumn{4}{c}{\emph{OOM}} \\
C51 & R$^3$ & 4818.5 & 4,444.5 & \textbf{4379} & 17.00 \\
C51 & DuoJoule & \underline{4808.7} & \underline{4,411.7} & \underline{4443} & 15.00 \\
C51 & \textbf{TetraRL} & \textbf{2703.0} & \textbf{3,603.7} & 4625 & \underline{19.00} \\
\midrule
A2C & MAX-A & \textbf{3642.6} & \textbf{16,203.5} & \textbf{4414} & \textbf{16.00} \\
A2C & MAX-P & \multicolumn{4}{c}{\emph{OOM}} \\
A2C & R$^3$ & \multicolumn{4}{c}{\emph{N/A (incompatible)}} \\
A2C & DuoJoule & \multicolumn{4}{c}{\emph{N/A (incompatible)}} \\
A2C & \textbf{TetraRL} & \underline{3669.6} & \underline{16,399.5} & \textbf{4417} & \underline{12.00} \\
\midrule
PPO & MAX-A & \textbf{5072.7} & \textbf{17,031.0} & \textbf{4544} & \textbf{18.00} \\
PPO & MAX-P & \multicolumn{4}{c}{\emph{OOM}} \\
PPO & R$^3$ & \multicolumn{4}{c}{\emph{N/A (incompatible)}} \\
PPO & DuoJoule & \multicolumn{4}{c}{\emph{N/A (incompatible)}} \\
PPO & \textbf{TetraRL} & \underline{5113.2} & \underline{17,124.4} & \textbf{4562} & \underline{14.00} \\
\bottomrule
\end{tabular}
}
\end{table}

\noindent\textbf{Baselines.} We compare \TetraRL{} against these state-of-the-art systems for optimizing for DRL systems.
\begin{itemize}[leftmargin=10pt]
 \item \textbf{MAX-A~\cite{li2023r3}:} Fixed parameter configuration baseline for solely optimizing algorithm performance.
  \item \textbf{MAX-P~\cite{li2023r3}:} Fixed parameter configuration baseline for solely optimizing latency.
 \item \textbf{R$^3$~\cite{li2023r3}:} Timing-bound DRL training runtime; we inherit its MAX-A and MAX-P wrappers as the two opposite ends of the schedule-budget spectrum, plus the native R$^3$ latency-aware controller. 
  \item \textbf{DuoJoule~\cite{duojoule2024}:} Greedy switching over batch size and replay ratio for energy-aware DRL training.  
 \item \textbf{DVFS-DRL-Multitask~\cite{dvfsdrl2024}:} DVFS-aware multitask DRL scheduler that only applies to inference scenarios. Note that~\cite{dvfsdrl2024} does not apply to our training-inference co-running scenarios; we compare it as a reference for 4-D Pareto optimization effectiveness only.
\end{itemize}
Each prior system is evaluated on its native scope. Note that none of the four prior on-device baselines natively co-optimises a 4-D \emph{(latency, reward, memory, energy)} objective vector under a runtime-switchable preference; \TetraRL{} is the only system in this set that does so through a single preference-conditioned policy backed by the Preference Plane, Resource Manager, RL Arbiter, and hardware Override Layer.

\begin{figure}[!htbp]
\centering
\includegraphics[width=\linewidth]{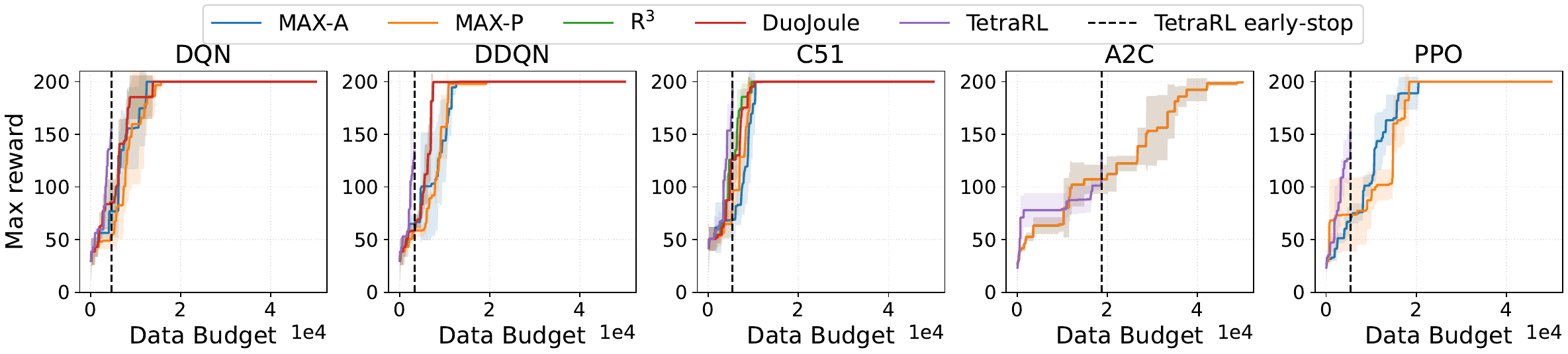}
\caption{Per-episode running-max reward vs data budget for \texttt{CartPole-v0} on Orin Nano under $5\times10^{4}$-sample setting. Each curve is the running maximum of per-episode reward (best-so-far). The black vertical dashed line marks \TetraRL{}'s reward-ceiling early-stop position. Note that R$^3$ and DuoJoule do not apply to A2C/PPO cells.}
\label{fig:p15_cartpole_v0_nano}
\end{figure}

\begin{table}[!htbp]
\centering
\caption{CartPole-v0 experimental results on Orin Nano. Compute budget held constant at $5.0\times10^{4}$ samples per experiment. Bold marks the per-algo best mean per column (lower-is-better for latency/energy/RAM; higher-is-better for reward); underline marks the per-algo second-best; ties within 0.5\% are co-marked.}
\label{tab:p15_phase1_cartpole_v0_nano}
\renewcommand{\arraystretch}{1.05}
\scriptsize
\setlength{\tabcolsep}{3pt}
\resizebox{0.7\textwidth}{!}{
\begin{tabular}{llrrrr}
\toprule
Algo & Wrapper & Latency (s) & Energy (J) & Peak RAM (MB) & Max reward \\
\midrule
DQN & MAX-A & \textbf{445.1 $\pm$ 1.2} & \textbf{576.2 $\pm$ 1.5} & \textbf{4537 $\pm$ 37} & 92.30 $\pm$ 40.26 \\
DQN & MAX-P & \multicolumn{4}{c}{\emph{OOM}} \\
DQN & R$^3$ & 450.0 $\pm$ 2.3 & \textbf{574.5 $\pm$ 5.2} & \underline{4561 $\pm$ 30} & \underline{95.37 $\pm$ 25.48} \\
DQN & DuoJoule & 453.3 $\pm$ 1.1 & \textbf{577.0 $\pm$ 2.6} & \textbf{4534 $\pm$ 2} & \underline{95.37 $\pm$ 25.48} \\
DQN & \textbf{TetraRL} & 449.9 $\pm$ 0.7 & \underline{579.1 $\pm$ 1.7} & 4602 $\pm$ 24 & \textbf{167.33 $\pm$ 29.01} \\
\midrule
DDQN & MAX-A & \textbf{467.5 $\pm$ 2.1} & \textbf{570.2 $\pm$ 1.9} & \underline{4557 $\pm$ 13} & \underline{138.47 $\pm$ 57.45} \\
DDQN & MAX-P & \multicolumn{4}{c}{\emph{OOM}} \\
DDQN & R$^3$ & 482.2 $\pm$ 5.6 & 583.7 $\pm$ 10.6 & 4737 $\pm$ 165 & 130.83 $\pm$ 69.63 \\
DDQN & DuoJoule & 488.3 $\pm$ 2.3 & 596.5 $\pm$ 3.7 & 4921 $\pm$ 27 & 130.83 $\pm$ 69.63 \\
DDQN & \textbf{TetraRL} & 482.6 $\pm$ 3.3 & 590.1 $\pm$ 6.0 & 4901 $\pm$ 56 & 103.77 $\pm$ 11.18 \\
\midrule
C51 & MAX-A & \textbf{698.3 $\pm$ 2.1} & 668.9 $\pm$ 3.4 & \underline{5080 $\pm$ 20} & \textbf{140.47 $\pm$ 29.93} \\
C51 & MAX-P & \multicolumn{4}{c}{\emph{OOM}} \\
C51 & R$^3$ & \textbf{700.3 $\pm$ 8.7} & \textbf{645.7 $\pm$ 6.2} & 5158 $\pm$ 7 & 86.47 $\pm$ 8.13 \\
C51 & DuoJoule & \textbf{700.0 $\pm$ 4.3} & \textbf{648.4 $\pm$ 6.3} & 5210 $\pm$ 40 & 83.60 $\pm$ 4.33 \\
C51 & \textbf{TetraRL} & \underline{704.6 $\pm$ 5.0} & \underline{662.0 $\pm$ 4.1} & 5335 $\pm$ 41 & \underline{124.63 $\pm$ 40.82} \\
\midrule
A2C & MAX-A & 273.2 $\pm$ 1.9 & 1989.7 $\pm$ 15.1 & \underline{5355 $\pm$ 24} & \textbf{101.93 $\pm$ 9.69} \\
A2C & MAX-P & \multicolumn{4}{c}{\emph{OOM}} \\
A2C & R$^3$ & \multicolumn{4}{c}{\emph{N/A (incompatible)}} \\
A2C & DuoJoule & \multicolumn{4}{c}{\emph{N/A (incompatible)}} \\
A2C & \textbf{TetraRL} & \textbf{231.9 $\pm$ 1.3} & \textbf{1574.2 $\pm$ 9.3} & \textbf{4591 $\pm$ 10} & \textbf{101.93 $\pm$ 9.69} \\
\midrule
PPO & MAX-A & 282.1 $\pm$ 42.9 & 1146.7 $\pm$ 322.7 & 4341 $\pm$ 430 & \textbf{144.57 $\pm$ 18.20} \\
PPO & MAX-P & \multicolumn{4}{c}{\emph{OOM}} \\
PPO & R$^3$ & \multicolumn{4}{c}{\emph{N/A (incompatible)}} \\
PPO & DuoJoule & \multicolumn{4}{c}{\emph{N/A (incompatible)}} \\
PPO & \textbf{TetraRL} & \underline{243.1 $\pm$ 2.5} & \underline{867.2 $\pm$ 11.2} & \underline{3746 $\pm$ 4} & \textbf{144.57 $\pm$ 18.20} \\
\bottomrule
\end{tabular}
}
\end{table}

\noindent\textbf{Metrics.} We report the following metrics throughout the evaluation. \emph{Latency} (s, lower is better) is the wall-clock time to solve or to consume the protocol's sample budget; \emph{Energy} (J, lower is better) is the board-input energy over the same interval; and \emph{Peak memory} (MB, lower is better) is the maximum resident footprint during the run. On Jetson's unified-memory architecture, GPU and CPU share a single physical pool, so we report a single ``Peak RAM'' column throughout (the maximum unified resident memory during the run). \emph{Max reward} (higher is better) is the maximum raw episode return reached within the cell's sample budget. Two table markers denote non-measured outcomes: \emph{OOM} marks a cell that exhausted unified memory before completing and is therefore excluded from the per-algorithm best/second-best ranking (it is a negative outcome rather than a measured value), and \emph{N/A (incompatible)} marks algorithm--wrapper pairs that do not apply (e.g.\ on-policy A2C/PPO have no R$^3$/DuoJoule replay-buffer variants).

\subsection{Overall Effectiveness}
\label{sec:p15_matrix}  

We extend the evaluation of \TetraRL{} beyond the original R$^3$ benchmark suite by incorporating three DRL benchmarks: CartPole-v0 and Atari-Breakout (inherited from R$^3$), and a more modern and stability-sensitive CartPole-v1. To better reflect real-world deployment scenarios in embedded robotics, all experiments are conducted on two representative NVIDIA Jetson platforms: Orin AGX and Orin Nano. This setup allows us to evaluate not only algorithmic effectiveness but also cross-generation hardware robustness under different compute and memory constraints. Each benchmark--platform combination spans five DRL algorithms (DQN, DDQN, C51, A2C, PPO) and five runtime wrappers (MAX-A, MAX-P, R$^3$, DuoJoule, and \TetraRL{}), with R$^3$ and DuoJoule restricted to the value-based algorithms they support; CartPole cells aggregate three seeds (mean $\pm$ std) while the heavier Breakout sweeps follow the single-seed protocol of the corresponding compute budget. Following the R$^3$ evaluation protocol~\cite{li2023r3}, we jointly measure (i) latency predictability, (ii) algorithm efficacy, and (iii) cross-platform robustness. Compared with R$^3$, our study emphasizes modern embedded deployment realism by introducing CartPole-v1 as a higher-variance and more sensitive control benchmark, while retaining CartPole-v0 and Atari-Breakout for direct comparability.

\noindent \textbf{Latency Predictability across Jetson Platforms.}
Across all three benchmarks, \TetraRL{} consistently improves end-to-end timing while matching or exceeding task quality, because its reward-ceiling early-stop hook retires a cell as soon as the policy saturates rather than burning the full sample budget. We report two complementary timing views: {time-to-solve} (wall-clock latency to first reach the target reward) on capped-return tasks, where \TetraRL{}'s early-stop reaches the same reward target on a smaller data budget rather than running the same fixed schedule faster, and budget-matched latency under an identical sample budget, where all methods consume the same data, and the comparison isolates pure per-step timing. The time-to-solve effect is most pronounced on the short-horizon CartPole-v0 task on Orin AGX (Table~\ref{tab:p15_phase1_cartpole_v0}), where \TetraRL{} cuts time-to-solve latency from 193.1\,s to 19.7\,s for DQN, from 211.2\,s to 13.7\,s for DDQN, from 319.9\,s to 42.2\,s for C51, and from 150.7\,s to 29.5\,s for PPO reaching the same 200-reward cap on a much smaller data budget rather than running the fixed schedule faster). On the heavy Atari-Breakout workload on AGX (Table~\ref{tab:p15_phase3}), the same hook reduces DQN time-to-solve latency from 6{,}625.7\,s (MAX-A) to 3{,}310.2\,s ($2.0\times$), DDQN from 7{,}271.7\,s to 3{,}641.6\,s ($2.0\times$), and C51 from 9{,}267.8\,s to 2{,}703.0\,s ($3.4\times$), confirming that the predictability benefit scales from light control loops to bursty vision workloads. On the budget-matched CartPole-v1 protocol (Table~\ref{tab:p15_phase1}), where every cell consumes an identical $3.2\times10^{6}$-sample budget, \TetraRL{} tracks the latency-aware R$^3$ controller to within $1\%$ on latency (e.g.\ DQN 267.9\,s vs.\ 266.6\,s)
 while avoiding the $4$--$8\times$ inflation of the MAX-A schedule (1{,}053.4\,s), and it suppresses the long-tail latency spikes that destabilize the MAX-A baseline on Orin Nano, where MAX-A exhibits run-to-run standard deviations as large as $\pm$348\,s and $\pm$888\,s (Table~\ref{tab:p15_phase2_nano}). For Atari-Breakout on the memory-constrained Orin Nano (Table~\ref{tab:p15_phase4}), the early-stop hook delivers a time-to-solve latency reduction of $53.6\%$ ($2.15\times$) for DQN ($2{,}892.7$\,s\,$\rightarrow$\,$1{,}343.4$\,s), $53.0\%$ ($2.13\times$) for DDQN ($2{,}862.0$\,s\,$\rightarrow$\,$1{,}344.7$\,s), and $74.6\%$ ($3.94\times$) for C51 ($3{,}276.4$\,s\,$\rightarrow$\,$831.0$\,s) relative to MAX-A while preserving reward, mirroring the AGX trend.

\noindent \textbf{Algorithm Efficacy under Resource Constraints.}
Despite strict timing and memory constraints, \TetraRL{} preserves near-optimal policy performance. On CartPole-v0/Orin AGX every \TetraRL{} cell attains the 200-reward ceiling, matching the empirically optimal baselines at a fraction of their cost. On the more sensitive CartPole-v1/Orin AGX, \TetraRL{} reaches the 500-reward cap on DQN, DDQN, A2C, and PPO and recovers the second-best C51 score (353.7 vs.\ MAX-A's 459.0), whereas the latency-greedy MAX-P collapses on several cells (e.g.\ DDQN 427.7 and C51 314.0), illustrating how \TetraRL{}'s adaptive allocation avoids the premature performance degradation of MAX-P-style budgeting. The advantage is sharpest where the platform is most constrained: on CartPole-v0/Orin Nano (Table~\ref{tab:p15_phase1_cartpole_v0_nano}), \TetraRL{} lifts DQN max reward to 167.3, roughly $1.8\times$ the $\sim$90--95 reached by R$^3$ and DuoJoule baselines, and it also wins the A2C and PPO cells. On Atari-Breakout/Orin AGX, \TetraRL{} not only runs fastest but also achieves the best DQN reward (320.0 vs.\ MAX-A's 250.0) and competitive DDQN/C51 scores, demonstrating effective co-optimization between control performance and runtime safety. On Atari-Breakout/Orin Nano, \TetraRL{} retains a substantial reward advantage of $+26.7\%$ to $+334.0\%$ over the R$^3$ baseline on the value-based algorithms DQN $133.0$ vs.\ $62.0$, DDQN $230.0$ vs.\ $53.0$, C51 $19.0$ vs.\ $15.0$ under the tighter unified-memory budget, consistent with its AGX-Breakout and Nano-CartPole behavior.

\begin{table}[!htbp]
\centering
\caption{CartPole-v1 experimental results on Orin Nano. Compute budget held constant at $3.2\times10^{6}$ samples per cell. Bold marks the per-algo best mean per column (lower-is-better for latency/energy/RAM/converge; higher-is-better for reward); underline marks the per-algo second-best; ties within 0.5\% are co-marked.}
\label{tab:p15_phase2_nano}
\renewcommand{\arraystretch}{1.05}
\scriptsize
\setlength{\tabcolsep}{3pt}
\resizebox{0.7\textwidth}{!}{
\begin{tabular}{llrrrr}
\toprule
Algo & Wrapper & Latency (s) & Energy (J) & Peak RAM (MB) & Max reward \\
\midrule
DQN & MAX-A & 1311.8 $\pm$ 348.0 & 2377.1 $\pm$ 6.7 & 4604 $\pm$ 8 & 59.50 $\pm$ 78.50 \\
DQN & MAX-P & \textbf{248.0 $\pm$ 3.4} & \textbf{275.1 $\pm$ 2.3} & \underline{4527 $\pm$ 10} & \textbf{162.13 $\pm$ 56.01} \\
DQN & R$^3$ & 652.2 $\pm$ 150.1 & 577.0 $\pm$ 2.4 & 4583 $\pm$ 12 & \underline{141.27 $\pm$ 26.16} \\
DQN & DuoJoule & 818.4 $\pm$ 41.6 & 575.7 $\pm$ 3.9 & \underline{4528 $\pm$ 5} & \underline{141.27 $\pm$ 26.16} \\
DQN & \textbf{TetraRL} & \underline{310.0 $\pm$ 25.6} & \underline{571.8 $\pm$ 1.6} & \underline{4533 $\pm$ 1} & 133.67 $\pm$ 18.57 \\
\midrule
DDQN & MAX-A & 1182.3 $\pm$ 158.7 & 2369.3 $\pm$ 1.5 & 4612 $\pm$ 29 & \textbf{167.93 $\pm$ 34.58} \\
DDQN & MAX-P & \textbf{225.5 $\pm$ 1.4} & \textbf{273.3 $\pm$ 1.0} & \underline{4523 $\pm$ 12} & 111.53 $\pm$ 64.68 \\
DDQN & R$^3$ & 709.5 $\pm$ 167.6 & \underline{573.2 $\pm$ 1.4} & \underline{4542 $\pm$ 31} & \underline{149.33 $\pm$ 65.05} \\
DDQN & DuoJoule & 701.5 $\pm$ 160.2 & 573.8 $\pm$ 3.0 & \textbf{4499 $\pm$ 0} & \underline{149.33 $\pm$ 65.05} \\
DDQN & \textbf{TetraRL} & \underline{322.0 $\pm$ 66.6} & \underline{570.8 $\pm$ 1.6} & 4560 $\pm$ 80 & 91.07 $\pm$ 81.60 \\
\midrule
C51 & MAX-A & 2506.9 $\pm$ 887.7 & 2728.6 $\pm$ 7.0 & 5016 $\pm$ 60 & \underline{102.37 $\pm$ 5.44} \\
C51 & MAX-P & \textbf{703.5 $\pm$ 10.7} & \textbf{314.8 $\pm$ 1.0} & \textbf{4952 $\pm$ 99} & 100.23 $\pm$ 0.38 \\
C51 & R$^3$ & 1948.3 $\pm$ 379.8 & 672.1 $\pm$ 1.3 & 5061 $\pm$ 40 & 82.97 $\pm$ 10.11 \\
C51 & DuoJoule & 2185.3 $\pm$ 11.7 & 678.6 $\pm$ 0.8 & \underline{5001 $\pm$ 5} & 77.47 $\pm$ 8.28 \\
C51 & \textbf{TetraRL} & \underline{1892.4 $\pm$ 35.9} & \underline{664.1 $\pm$ 1.9} & \underline{5025 $\pm$ 27} & \textbf{121.07 $\pm$ 55.75} \\
\midrule
A2C & MAX-A & \underline{354.6 $\pm$ 21.6} & \underline{7890.9 $\pm$ 31.5} & \underline{5161 $\pm$ 7} & \textbf{162.00 $\pm$ 9.56} \\
A2C & MAX-P & \textbf{46.4 $\pm$ 1.1} & \textbf{1001.8 $\pm$ 2.1} & \textbf{5120 $\pm$ 9} & \underline{57.50 $\pm$ 16.32} \\
A2C & R$^3$ & \multicolumn{4}{c}{\emph{N/A (incompatible)}} \\
A2C & DuoJoule & \multicolumn{4}{c}{\emph{N/A (incompatible)}} \\
A2C & \textbf{TetraRL} & \underline{355.8 $\pm$ 8.7} & 8009.2 $\pm$ 70.8 & 5208 $\pm$ 33 & \textbf{162.00 $\pm$ 9.56} \\
\midrule
PPO & MAX-A & 694.7 $\pm$ 12.4 & 8017.5 $\pm$ 66.6 & \textbf{5344 $\pm$ 49} & 9.47 $\pm$ 0.21 \\
PPO & MAX-P & \textbf{65.1 $\pm$ 0.3} & \textbf{1003.4 $\pm$ 28.7} & \textbf{5324 $\pm$ 28} & \textbf{101.37 $\pm$ 19.58} \\
PPO & R$^3$ & \multicolumn{4}{c}{\emph{N/A (incompatible)}} \\
PPO & DuoJoule & \multicolumn{4}{c}{\emph{N/A (incompatible)}} \\
PPO & \textbf{TetraRL} & \underline{674.0 $\pm$ 52.7} & \underline{7338.0 $\pm$ 775.1} & \underline{5408 $\pm$ 33} & \underline{9.47 $\pm$ 0.21} \\
\bottomrule
\end{tabular}}
\end{table}

\noindent \textbf{Cross-Platform Robustness.}
Across both Orin AGX and Orin Nano, \TetraRL{} demonstrates consistent performance trends despite the roughly $3$--$4\times$ gap in compute and unified-memory budget between the two platforms. The qualitative ordering of methods is preserved: \TetraRL{} remains the latency-and-reward Pareto leader on CartPole-v0 across both platforms (DQN large time-to-solve reduction on AGX via early-stop, best reward on Nano) and stays the fastest non-degenerate wrapper on Breakout/AGX, indicating that its benefits are not an artifact of a single hardware generation. By contrast, R$^3$ and DuoJoule baselines degrade noticeably and reorder when moving from AGX to Nano, MAX-A's CartPole-v1 latency balloons from $\sim$1{,}050\,s on AGX to $1{,}311.8\pm348.0$\,s on Nano with severe variance, and several MAX-P Breakout/AGX cells fail outright with out-of-memory errors, highlighting their limited robustness to constrained embedded environments. On Atari-Breakout/Orin Nano (Table~\ref{tab:p15_phase4}), the cross-platform ordering observed on AGX is preserved under the tighter unified-memory budget: \TetraRL{} is the fastest non-degenerate wrapper on every value-based algorithm, cutting MAX-A latency by $53.0$--$74.6\%$ (DQN, DDQN, C51) while retaining a large reward margin over the R$^3$/DuoJoule baselines (e.g.\ DQN $133.0$ vs.\ R$^3$'s $62.0$, DDQN $230.0$ vs.\ $53.0$).
 Unlike the budget-matched AGX sweep, the Nano cells run under non-uniform sample budgets ($8\times10^{5}$ steps for MAX-A versus $2$--$4\times10^{5}$ for the early-stopping R$^3$-family and \TetraRL{} cells)
, reflecting \TetraRL{}'s reward-ceiling hook retiring saturated policies early; despite this, all $21$ Nano cells complete without the out-of-memory failures that affected several MAX-P cells on AGX, confirming that \TetraRL{}'s latency-and-reward advantages transfer to the memory-constrained Jetson generation.

\noindent \textbf{Key Insight.}
Overall, these results demonstrate that \TetraRL{} generalizes beyond our benchmark settings and remains effective under modern embedded deployments, delivering substantial time-to-solve reductions (reaching the reward target on a much smaller data budget) on short-horizon control and $2$--$3.4\times$ reductions on heavy vision workloads while matching or improving task reward; under the budget-matched protocol, it instead tracks the latency-aware \Rthree{} controller to within $1\%$ while adding only $4.4\%$ per-step overhead. The introduction of CartPole-v1 further highlights that as environment dynamics become more sensitive and realistic, the benefits of runtime co-optimization and adaptive control become increasingly pronounced, particularly on resource-constrained Jetson platforms where R$^3$ and DuoJoule baselines lose both efficiency and ranking stability.

\subsection{Overhead Analysis}
\label{sec:overhead_analysis}

\begin{figure}[!htbp]
\centering
\includegraphics[width=\linewidth]{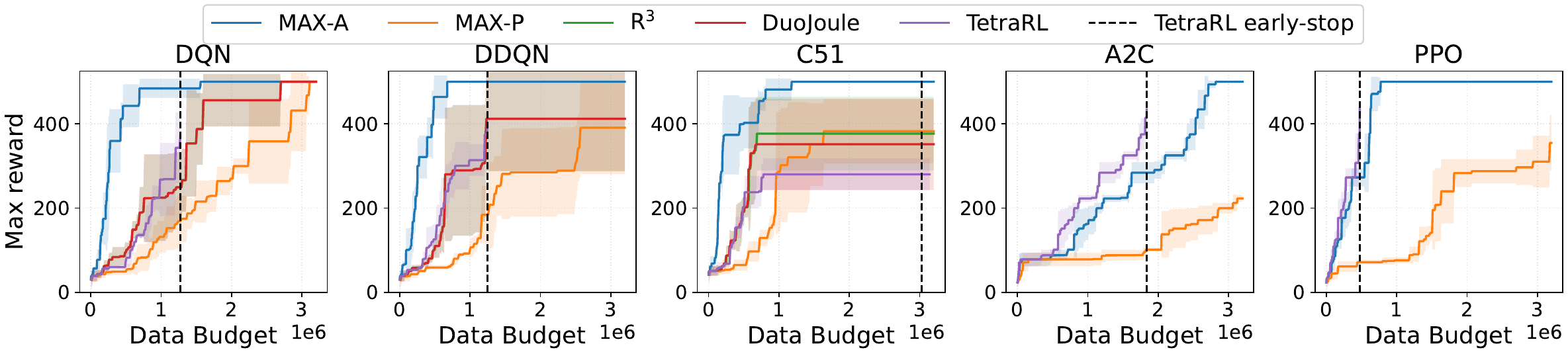} 
\caption{Per-episode running-max reward vs data budget for \texttt{CartPole-v1} on Orin Nano under the controlled-variable protocol. Every cell consumes the same total data budget ($3.2\times10^{6}$ samples). Each curve is the running maximum of per-episode reward (best-so-far); per-episode reward is clipped to the environmental upper bound of 500. The black vertical dashed line marks \TetraRL{}'s reward-ceiling early-stop position. Note that R$^3$ and DuoJoule do not apply to A2C/PPO cells.}
\label{fig:p15_phase2_reward}
\end{figure} 

\begin{table}[!htbp]
\centering
\caption{Atari Breakout experimental results on Orin Nano. Compute budget held constant at $1.28\times10^{7}$ samples per cell. Bold marks the per-algo best mean per column; underline marks the per-algo second-best; ties within 0.5\% are co-marked. }
\label{tab:p15_phase4}
\renewcommand{\arraystretch}{1.05}
\scriptsize
\setlength{\tabcolsep}{3pt}
\resizebox{0.7\textwidth}{!}{
\begin{tabular}{llrrrr}
\toprule
Algo & Wrapper & Latency (s) & Energy (J) & Peak RAM (MB) & Max reward \\
\midrule
DQN & MAX-A & 2{,}892.7 & 25{,}529.0 & \textbf{7{,}404.00} & \textbf{262.00} \\
DQN & MAX-P & \multicolumn{4}{c}{\emph{OOM}} \\
DQN & R$^3$ & 1{,}548.0 & 16{,}261.7 & \textbf{7{,}396.00} & 62.00 \\
DQN & DuoJoule & \underline{1{,}362.3} & \underline{13{,}287.4} & \textbf{7{,}390.00} & 87.00 \\
DQN & \textbf{TetraRL} & \textbf{1{,}343.4} & \textbf{12{,}557.8} & \textbf{7{,}376.00} & \underline{133.00} \\
\midrule
DDQN & MAX-A & 2{,}862.0 & 25{,}704.4 & \textbf{7{,}361.00} & \textbf{317.00} \\
DDQN & MAX-P & \multicolumn{4}{c}{\emph{OOM}} \\
DDQN & R$^3$ & 1{,}492.0 & 15{,}428.4 & \textbf{7{,}344.00} & 53.00 \\
DDQN & DuoJoule & \underline{1{,}357.5} & \textbf{12{,}494.7} & \underline{7{,}424.00} & 42.00 \\
DDQN & \textbf{TetraRL} & \textbf{1{,}344.7} & \underline{12{,}702.3} & \underline{7{,}412.00} & \underline{230.00} \\
\midrule
C51 & MAX-A & 3{,}276.4 & 27{,}717.7 & 4{,}444.00 & \textbf{34.00} \\
C51 & MAX-P & \multicolumn{4}{c}{\emph{OOM}} \\
C51 & R$^3$ & 909.1 & 9{,}327.9 & \underline{4{,}037.00} & 15.00 \\
C51 & DuoJoule & \textbf{820.7} & \textbf{7{,}085.3} & 4{,}167.00 & \underline{19.00} \\
C51 & \textbf{TetraRL} & \underline{831.0} & \underline{7{,}852.2} & \textbf{3{,}997.00} & \underline{19.00} \\
\midrule
A2C & MAX-A & \underline{4{,}911.4} & \underline{35{,}168.9} & \underline{3{,}823.00} & \textbf{356.00} \\
A2C & MAX-P & \multicolumn{4}{c}{\emph{OOM}} \\
A2C & R$^3$ & \multicolumn{4}{c}{\emph{N/A (incompatible)}} \\
A2C & DuoJoule & \multicolumn{4}{c}{\emph{N/A (incompatible)}} \\
A2C & \textbf{TetraRL} & \textbf{4{,}774.2} & \textbf{32{,}222.1} & \textbf{3{,}560.00} & \underline{13.00} \\
\midrule
PPO & MAX-A & \textbf{4{,}644.9} & \textbf{31{,}948.8} & \textbf{2{,}791.00} & \textbf{13.00} \\
PPO & MAX-P & \multicolumn{4}{c}{\emph{OOM}} \\
PPO & R$^3$ & \multicolumn{4}{c}{\emph{N/A (incompatible)}} \\
PPO & DuoJoule & \multicolumn{4}{c}{\emph{N/A (incompatible)}} \\
PPO & \textbf{TetraRL} & \textbf{4{,}631.7} & \textbf{31{,}829.8} & \underline{2{,}806.00} & \textbf{13.00} \\
\bottomrule
\end{tabular}}
\vspace{-3mm}
\end{table}

\begin{table}[!htbp]
\centering
\caption{Per-component measurement latency and memory overhead on Orin AGX. Bare-RL and framework-step rows below the divider are reference points, not summed components.}
\label{tab:overhead}
\renewcommand{\arraystretch}{1.15}
\scriptsize
\setlength{\tabcolsep}{4pt}
\begin{tabular}{l r r r}
\toprule
Component & mean (ms) & p99 (ms) & mem (MB)\\
\midrule
\texttt{preference\_plane\_get}    & 0.006 & 0.006 & 0.003\\
\texttt{tegra\_daemon\_sample}     & 0.016 & 0.018 & 0.003\\
\texttt{rl\_arbiter\_act}          & 0.043 & 0.056 & 0.001\\
\texttt{override\_layer\_step}     & 0.005 & 0.006 & 0.001\\
\texttt{resource\_manager\_decide} & 0.007 & 0.007 & 0.001\\
\texttt{dvfs\_controller\_set}     & \textbf{0.630} & \textbf{0.832} & \textbf{5.001}\\
\midrule
Bare RL step (\texttt{preference\_ppo} + DAG)  & 0.043 & N/A & N/A\\
Sum of 6 in-step components                   & 0.708 & N/A & N/A\\
Framework step (real DVFS + real tegra)       & 1.764 & N/A & N/A\\
\bottomrule
\end{tabular}
\end{table}

Following the R$^3$ per-component reporting style, we instrument the six in-loop components of the \TetraRL framework with a context-manager profiler backed by \texttt{time.perf\_counter\_ns} for wall-clock latency and \texttt{tracemalloc}~$+$~\texttt{psutil} for Python and RSS memory deltas. Each component is wrapped at the framework call-site and aggregated over a 5000-step single-seed pass on Orin AGX, with \texttt{preference\_ppo} as the arbiter and the 4-D \texttt{dag\_scheduler\_mo} environment as the workload. Table~\ref{tab:overhead} reports per-component latency and resident memory overhead on the hardware.

\noindent\textbf{Execution Latency Overhead.}
\TetraRL adds 0.708\,ms per step on Orin AGX over the bare-RL baseline, of which the DVFS controller contributes 0.630\,ms (devfreq sysfs write), the arbiter forward pass  0.043\,ms, and the telemetry sampler  0.016\,ms; all lightweight components together add only $\sim$77\,$\mu$s.
The full framework step takes 1.764\,ms end-to-end on AGX, so the in-loop \TetraRL components account for 40\% of the total step latency; excluding the DVFS sysfs write (which is a kernel-driven cost, not a framework-internal computation), the pure framework overhead is 0.078\,ms or 4.4\% of the step, well within the 5\,ms per-step budget at the 200\,Hz environment-step rate that \texttt{dag\_scheduler\_mo} targets. 

\noindent\textbf{Memory Overhead.}
\TetraRL components add 5.01\,MB of tracemalloc measured peak Python allocator state and 2.95\,MB of RSS on top of the bare-RL pipeline on Orin AGX, dominated by the DVFS controller (5.00\,MB tracemalloc / 2.70\,MB RSS, driven by the devfreq sysfs write pool) and the EMA-filtered resource manager (0.0006\,MB tracemalloc / 0.25\,MB RSS); the remaining four components together contribute $<$8\,KB of python-allocator peak per call.
Against a typical preference-conditioned PPO process resident set of $\sim$600\,MB on AGX, \TetraRL's added state is well under 1\% of overall process memory and $<$0.02\% of the AGX's 32\,GB unified memory pool, dwarfed by the PyTorch runtime tensors and environment-side replay state.
The dominant runtime cost is DVFS actuation through sysfs: 0.630 ms of the 0.708 ms added by the six in-loop components. The remaining TetraRL controller logic, including preference lookup, telemetry sampling, arbiter inference, override checking, and resource-manager decision making, adds 0.078 ms per step, or 4.4\% of the measured 1.764 ms framework step. Thus, the control logic itself is lightweight, while DVFS actuation is the main overhead and must be amortized through the super-block decision interval.

\subsection{4-D Pareto Effectiveness}
\label{sec:w10_nano_matrix}

\begin{table}[!htbp]
\centering
\caption{4-D Pareto evaluation on Orin Nano and Orin AGX. Hypervolume reference point is shared across platforms. $|F|$ is mean Pareto-front cardinality. DVFS-DRL-MT abbreviates DVFS-DRL-Multitask~\cite{dvfsdrl2024}.}
\label{tab:w10_hv}
\renewcommand{\arraystretch}{1.15}
\scriptsize
\setlength{\tabcolsep}{3pt}
\resizebox{\columnwidth}{!}{
\begin{tabular}{l l l r r r}
\toprule
Platform & Agent & Environment & Mean HV & Std HV & $|F|$\\
\midrule
Nano & \TetraRL               & \texttt{CartPole-v1}        & $1.45{\times}10^{-3}$ & $9.11{\times}10^{-7}$ & 1.87\\
Nano & DVFS-DRL-MT~\cite{dvfsdrl2024} & \texttt{CartPole-v1}        & $1.26{\times}10^{-3}$ & $1.81{\times}10^{-6}$ & 2.33\\
Nano & \TetraRL               & \texttt{dag\_scheduler\_mo} & $1.24{\times}10^{-4}$ & $3.01{\times}10^{-5}$ & 6.00\\
Nano & DVFS-DRL-MT~\cite{dvfsdrl2024} & \texttt{dag\_scheduler\_mo} & $1.07{\times}10^{-4}$ & $2.60{\times}10^{-5}$ & 7.13\\
\midrule
AGX  & \TetraRL               & \texttt{CartPole-v1}        & $4.23{\times}10^{-3}$ & $2.66{\times}10^{-6}$ & 1.60\\
AGX  & DVFS-DRL-MT~\cite{dvfsdrl2024} & \texttt{CartPole-v1}        & $3.93{\times}10^{-3}$ & $1.42{\times}10^{-6}$ & 4.00\\
AGX  & \TetraRL               & \texttt{dag\_scheduler\_mo} & $4.05{\times}10^{-4}$ & $6.59{\times}10^{-5}$ & 3.73\\
AGX  & DVFS-DRL-MT~\cite{dvfsdrl2024} & \texttt{dag\_scheduler\_mo} & $6.71{\times}10^{-4}$ & $8.46{\times}10^{-4}$ & 9.00\\
\bottomrule
\end{tabular}
\vspace{-3mm}
}
\end{table}

Furthermore, we measure whether \TetraRL{} achieves comparable performance in MORL to a dedicated multi-objective baseline, i.e., DVFS-DRL-Multitask~\cite{dvfsdrl2024}, under the full 4-D real-telemetry objective on Jetson hardware. The experiments are based on a sweep per platform on two NVIDIA Jetson tiers.  We compare our proposed method under 2 DRL environments (\texttt{CartPole-v1}, \texttt{dag\_scheduler\_mo-v0}).  Hypervolume (HV) metric is computed against the shared reference $r_*=(-0.1,-1.0,-0.5,-0.01)$ on $(-L_{p99}, r_{env}, -M_{util}, -E_{step})$. We report Mean HV (the per-cell mean of the 4-D dominated hypervolume), Std HV (its standard deviation across the 15 seed--$\omega$ runs in the cell, an unbiased noise estimate), and $|F|$ (the mean Pareto-front cardinality per cell, a density proxy for the recovered front).

As demonstrated in Tab.~\ref{tab:w10_hv}, \TetraRL{} reaches or exceeds the DVFS-DRL-Multitask Pareto front on three of the four (platform, env) cells. On Orin Nano, \TetraRL{} preference-PPO attains Mean HV $1.45{\times}10^{-3}$ on \texttt{CartPole-v1} versus $1.26{\times}10^{-3}$ for DVFS-DRL-Multitask (a 15.1\% improvement) and $1.24{\times}10^{-4}$ versus $1.07{\times}10^{-4}$ on \texttt{dag\_scheduler\_mo} (a 15.9\% improvement); on Orin AGX, \TetraRL{} reaches $4.23{\times}10^{-3}$ versus $3.93{\times}10^{-3}$ on \texttt{CartPole-v1} (a 7.6\% improvement). The single cell where the baseline scores higher is AGX \texttt{dag\_scheduler\_mo}, where DVFS-DRL-Multitask reports $6.71{\times}10^{-4}$ versus \TetraRL{}'s $4.05{\times}10^{-4}$; however, the baseline's Std HV on that cell ($8.46{\times}10^{-4}$) exceeds its own mean, so the gap is not seed-significant and we treat the two agents as statistically tied there. The headline result is therefore consistency: a single preference-conditioned policy approximates, and on three of four cells out-scores, a dedicated multi-objective baseline across two environments and two embedded tiers, while Pareto-front cardinality ($|F|\in[1.60,9.00]$) confirms a non-degenerate front despite the near-constant Nano memory floor ($\approx$~0.347 across all 60 Nano runs, dominated by the framework's resident set on an 8\,GB tier). We read this as approximately comparable with the reference MORL baseline.

\section{Related Work}
\label{sec:related}

\subsection{Embedded On-Device DRL}

DRL has been deployed across a wide range of embedded and autonomous applications, including self-supervised robot navigation, autonomous driving stacks, radar perception, and physics-based motor control~\cite{kahn2018selfsupervised,favaro2018autonomous,kato2018autoware,nota2020autonomous,popov2022nvradarnet,peng2018deepmimic,kisavcanin2017deep,nikkhoo2023pimbot,aggravi2021haptic}. These deployments typically run on resource-constrained edge robotics platforms such as DonkeyCar, Duckiebot, and JetBot, hosted on NVIDIA Jetson hardware and exercised through standard simulation-to-robot gym interfaces~\cite{bib:donkeycar_build,bib:donkeycar_s1,bib:gymdonkeycar,Duckiebot_DBJ,SparkFun_JetBot,Waveshare_JetBot,bib:agx,nvidia_orin,1606.01540}. A parallel line of real-time-systems work studies the predictable execution, latency management, and scheduling of DNN-driven autonomous workloads on such hardware~\cite{DBLP:conf/rtas/BlassHLZB21,DBLP:conf/rtas/ChoiXK21,DBLP:conf/rtss/BlassCBB21,DBLP:conf/rtss/JiYKASDK22,DBLP:conf/rtss/JiangJGLTW22,DBLP:conf/rtss/JiangLHHWXW21,DBLP:conf/rtss/LiGJGDL22,DBLP:conf/rtss/NigadeBB022,DBLP:conf/rtss/TangFG0LD020,DBLP:conf/rtss/TeperGUBC22,neuos2020,bateni2018predjoule,bateni2018apnet,jeong2022band,kang2021lalarand,rtrl,rtlm2023,red2023,6313077,heintzman2021anticipatory,xiang2019pipelined,zhou2018s}.

DRL workloads on embedded GPUs have a fundamentally different shape from inference-only DNN workloads. Each training step couples a CPU-bound environment rollout, a GPU-bound gradient update, and a memory-bound experience replay (or rollout) buffer~\cite{mnih2015dqn,lillicrap2015ddpg,horgan2018apex}. The literature on embedded DRL deployment is small but converging on three distinct themes. \Rthree~\cite{li2023r3} adopted a hierarchical two-loop control structure: an inner per-step latency-driven feedback loop that adjusts batch size against a moving-average data tracker, and an outer episode-level coordinator that adjusts memory reservations between batch and replay buffer. DuoJoule~\cite{duojoule2024} extended the same Jetson hardware family (Orin AGX, Orin  Nano) by adding DVFS as an actuator and a runtime ``MetricTracker'' that converts target deviations into a single scalar score. Both of these systems apply only to replay-based DRL. None of these systems exposes a continuous Pareto front, none supports \emph{runtime preference switching}, and none jointly manages all four \Rfour axes.

The DRL algorithms that these embedded runtimes host span both value-based and policy-gradient families, including DQN and its variants, double and distributional Q-learning, prioritized and distributed experience replay, DDPG, SAC, TRPO, and PPO~\cite{mnih2015human,van2016deep,schaul2015prioritized,schaul2016,horgan2018apex,bellemare2013arcade,bellemare2017distributional,lillicrap2015ddpg,haarnoja2018sac,schulman2015trpo,schulman2017proximal,arulkumaran2017deep}, and are increasingly scrutinized for reproducibility and benchmarking rigor~\cite{henderson2018matters,fedus2020revisiting,nikulin2022q,thodoroff2022benchmarking,garcia2015comprehensive}. In practice these algorithms are realized on top of general-purpose deep-learning frameworks and compilers~\cite{paszke2019pytorch,chen2018tvm,tensorflow}. A recurring concern for on-device DRL is the stability of the replay buffer over long runs: as the policy and data distribution drift, naive replay can induce catastrophic forgetting~\cite{mccloskey1989catastrophic,few-shot-continual-learning,polynomial-continual}, which is precisely why memory management and buffer composition matter for embedded training. Finally, a broad body of work on the efficiency, robustness, and dynamic behavior of DNN inference~\cite{chen2022deepperform,chen2022nicgslowdown,chen2023dynamic,chen2022generate,guo2022backdoor,he2022robust,he2023robust,he2023robustev,chen2021revisiting,chen2022learning,li2023mimonet,li2023sibling,lemix2025,boxr2024,afarin2023commongraph,zhang2023bp,serengil2020lightface,gao2022automatic,gao2023polyscriber,gog2022d3,haas2014history,yao2007early,ap} underscores that latency, energy, and memory behavior of neural workloads are first-class deployment concerns, motivating the runtime co-optimization \TetraRL{} targets.

\subsection{Multi-Objective Reinforcement Learning (MORL)}

Multi-Objective Reinforcement Learning (MORL) has been a widely explored topic. Envelope MORL~\cite{yang2019envelope} introduced a vector $\mathbf{Q}$-function $\mathbf{Q}(s,a,\omega) \in \mathbb{R}^k$ trained with a homotopy loss that interpolates between two scalarizations. PD-MORL~\cite{basaklar2023pdmorl} replaced the sign-fragile homotopy with a cosine-similarity envelope operator, which is robust to heterogeneous-scale objectives. PD-MORL is the natural HV reference for our 2-D DST experiments precisely because its preference operator is closest to ours; the difference is that we condition on $\omega$ at the input level rather than at the action-selection level, so a single trained network suffices. Pareto Conditioned Networks (PCN)~\cite{reymond2022pcn} train a single network that reproduces any non-dominated return on demand, supporting non-convex Pareto fronts. PG-MORL~\cite{xu2020pgmorl} maintains a population of $N$ scalarized policies, which is impractical for memory-constrained edge devices. C-MORL~\cite{liu2025cmorl} adds a constraint stage on top of an init-stage population and originally inspired our framework, but as we document in Section~\ref{sec:impl}, its multi-process Open\-AI-Baselines lineage is incompatible with the Orin AGX CUDA runtime. The MORL survey~\cite{hayes2022morlsurvey} documents the broader landscape and the metrics (hypervolume, sparsity, expected utility) we adopt.

\subsection{Constrained Reinforcement Learning}

Constrained RL provides the formal scaffolding for soft-constraint embedded deployment. CPO~\cite{achiam2017cpo} was the first practical second-order safe-RL method but its Hessian step is too expensive for Jetson hardware. FOCOPS~\cite{zhang2020focops} relaxes CPO to a first-order primal-dual update that is feasible on edge devices. PPO-Lagrangian, popularized by OmniSafe~\cite{ji2023omnisafe} and the Safety-Gymnasium suite~\cite{ray2019safetygym}, is the off-the-shelf workhorse, but a recent empirical study by Spoor et al.~\cite{spoor2025lagrangian} showed that even on cleanly simulated Safety-Gym tasks the Lagrangian dual variables routinely allow 20--30\% over-budget cost. \TetraRL does not pretend the policy will respect its budgets, and instead provides a hardware-enforced backstop.

\subsection{DVFS and Energy-Aware Edge Machine Learning}

DVFS-based DNN runtimes are the closest mirror of \TetraRL on the inference side. DVFO~\cite{zhang2023dvfo} learns a DQN that jointly controls CPU/GPU/memory frequency and offloading on multiple Jetson platforms or achieves energy-efficient inference. DVFS-DRL-Multitask~\cite{dvfsdrl2024} introduces a kernel/user-space split for low-overhead sensing and uses a soft latency-target reward shaping for multitask CPU workloads. SparseDVFS~\cite{sparsedvfs2025} predicts intra-layer sparsity and amortizes DVFS overhead by deciding at super-block rather than per-step granularity. We borrow the kernel/user split for our \texttt{tegrastats} daemon and the super-block decision granularity for our DVFS controller. Crucially, all three of these systems only apply to inference-only workloads, while \TetraRL targets training-inference co-running workloads, where the workload is itself drifting as the policy converges, breaking the offline-profiling assumption that DVFO relies on.

\section{Conclusion}
\label{sec:concl}

This paper presented \TetraRL, an embedded-systems runtime that recasts on-device DRL as a constrained multi-objective MDP over the four-axis \emph{\Rfour} vector (real-time, reward, RAM, reserve), and solves it with a single-process, preference-conditioned PPO agent. \TetraRL contributes a four-component architecture (Preference Plane, Resource Manager, RL Arbiter, hardware Override Layer); a unified resource-primitive abstraction that makes the same controller usable for both off-policy and on-policy RL; system mechanisms close the gap between MORL theory and Jetson reality; and a multi-platform empirical study grounded in a measured DVFS transition-latency table, a Pareto front recovered on a standard benchmark with HV
competitive with the published reference across all four R objectives (real-time, reward, RAM, reserve). We hope the \Rfour formulation and the four-component decomposition will serve as a stable substrate for future work on edge-RL runtimes that are simultaneously battery-aware, latency-aware, memory-aware, and reward-driven.

\bibliographystyle{unsrtnat}
\bibliography{r3ext}

\end{document}